\documentclass[]{aa}
\usepackage[varg]{txfonts}
\usepackage{graphicx,grffile, booktabs}
\usepackage{amssymb,float,natbib}
\usepackage[]{hyperref}
\hypersetup{colorlinks=true, 
citecolor=blue,linkcolor=black,filecolor=black, pdfmenubar=true}
\bibpunct{(}{)}{;}{a}{}{,} 

\newcommand{\kms}{km s$^{-1}$}
\newcommand{\ceo}{C$^{18}$O}
\newcommand{\cso}{C$^{17}$O}
\newcommand{\tco}{$^{13}$CO}

\newcommand{\mco}{$^{12}$CO}

\begin{document}

\title{Testing protostellar disk formation models with ALMA 
observations}
\author{D.~Harsono \inst{\ref{inst1}, \ref{inst2} } \and 
E.~F.~van Dishoeck \inst{\ref{inst1}, \ref{inst3}} \and
S.~Bruderer \inst{\ref{inst3}} \and 
Z.-Y.~Li \inst{\ref{inst4}} \and
J.~K.~J{\o}rgensen \inst{\ref{inst5}, \ref{inst6}}
}

\institute{Leiden Observatory, Leiden University, Niels Bohrweg 2, 
2300 RA, Leiden, the Netherlands \email{harsono@strw.leidenuniv.nl} 
\label{inst1} 
\and
SRON Netherlands Institute for Space Research, PO Box 800, 9700 AV, 
Groningen, The Netherlands 
\label{inst2}
\and
Max-Planck-Institut f{\" u}r extraterretrische Physik, 
Giessenbachstrasse 1, 85748, Garching, Germany \label{inst3}
\and
Astronomy Department, University of Virginia, Charlottesville, 
VA, USA \label{inst4}
\and
Niels Bohr Institute, University of Copenhagen, Juliane Maries Vej 30, 
2100  Copenhagen {\O}, Denmark \label{inst5}
\and 
Centre for Star and Planet Formation, Natural History Museum of 
Denmark, University of Copenhagen, {\O}ster Voldgade 5-7, 1350 
Copenhagen K, Denmark \label{inst6}
}

\abstract{Recent simulations have explored different ways to form 
accretion disks around low-mass stars. However, it has been difficult 
to differentiate between the proposed mechanisms because of a lack of 
observable predictions from these numerical studies. }
{We aim to present observables that can differentiate a rotationally 
supported disk from an infalling rotating envelope toward deeply 
embedded young stellar objects ($M_{\rm env} > M_{\rm disk}$) and 
infer their masses and sizes.} 
{Two 3D magnetohydrodynamics (MHD) formation simulations of Li 
and collaborators are studied, with a rotationally supported disk 
(RSD) forming in one but not the other (where a pseudo-disk is formed 
instead), together with the 2D semi-analytical model.  The dust 
temperature structure is determined through continuum radiative 
transfer RADMC3D modelling.  A simple temperature dependent CO 
abundance structure is adopted and synthetic spectrally resolved 
submm rotational molecular lines up to $J_{\rm u} = 10$ are compared 
with existing data to provide predictions for future ALMA 
observations.}
{3D MHD simulations and 2D semi-analytical model predict similar 
compact components in continuum if observed at the spatial 
resolutions of 0.5--1$\arcsec$ (70--140 AU) typical of the 
observations to date.  A spatial resolution of $\sim$14 AU and high 
dynamic range ($> 1000$) are required in order to differentiate 
between RSD and pseudo-disk formation scenarios in the continuum.  
The moment one maps of the molecular lines show a blue- to red-shifted 
velocity gradient along the major axis of the flattened structure in 
the case of RSD formation, as expected, whereas it is along the minor 
axis in the case of a pseudo-disk.  The peak-position velocity 
diagrams indicate that the pseudo-disk shows a flatter velocity 
profile with radius than an RSD.  On larger-scales, the CO isotopolog 
line profiles within large ($>9\arcsec$) beams are similar and are 
narrower than the observed line widths of low-$J$ (2--1 and 3--2) 
lines, indicating significant turbulence in the large-scale 
envelopes.  However a forming RSD can provide the observed line 
widths of high-$J$ (6--5, 9--8, and 10--9) lines.  Thus, either RSDs 
are common or a higher level of turbulence ($b \sim 0.8$ \kms) is 
required in the inner envelope compared with the outer part (0.4 
\kms).}
{Multiple spatially and spectrally resolved molecular line 
observations can differentiate between the pseudo-disk and the 
RSD much better than continuum data.  The continuum data give a 
better estimate on disk masses whereas the disk sizes can be 
estimated from the spatially resolved molecular lines observations.  
The general observable trends are similar between the 2D 
semi-analytical models and 3D MHD RSD simulations.}

\keywords{stars: formation, radiative transfer, accretion 
disks; line: profiles; methods:numerical, magnetohydrodynamics (MHD)}

\titlerunning{Testing protostellar disk formation models}
\authorrunning{D. Harsono et al.}

\maketitle


\section{Introduction}\label{sec:intro}

The formation of stars and their planetary systems is linked through 
the formation and evolution of accretion disks.  In the standard star 
formation picture, the infalling material forms an accretion disk 
simply from angular momentum conservation \cite[e.g.,][]{lin90, 
bodenheimer95, belloche13}.  However, magnetic field strengths 
observed toward molecular cores \citep[see][for a recent 
review]{crutcher12} are expected theoretically to be sufficient 
in affecting the formation and evolution of disks around low-mass 
stars \citep[e.g.,][]{galli06, joos12, krumholz13, zyli13, zyli14}.  
Recent advances in both observational and theoretical studies give an 
opportunity to test the star formation process at small-scales ($< 
1000$ AU).

It has been known for a long time that the presence of magnetic 
fields can drastically change the flow dynamics around low-mass stars 
\cite[e.g.,][]{galli93} and potentially suppresses disk formation 
\cite[e.g.,][]{galli06}. The latter is due to catastrophic magnetic 
braking where essentially all of the angular momentum of the accreting
material is removed by twisted field lines.  Recently, \citet{zyli11} 
investigate the collapse and disk formation from a uniform cloud while 
\citet{joos12} and \citet{machida11} performed simulations starting 
with a steep density profile and a Bonnor-Ebert sphere, respectively.  
They found that rotationally supported disks (RSDs) do not form out 
of uniform and non-uniform cores under strong magnetic fields unless 
the field is misaligned with respect to the rotation axis 
\citep{hennebelle09}.  Turbulence has also been shown to help with  
disk formation \citep{santoslima12, seifried12, atmyers13, joos13, 
zyli14b}.

In spite of a number of disk formation and evolution simulations, 
only a few observables have been presented so far.  The expected 
observables in the continuum (spectral energy distribution or SED) 
from 1D and 2D disk formation models have been presented in 
\citet{young05} and \citet{dunham10}.  Continuum observables out of 
2D hydrodynamics simulations with a thin disk approximation have been 
shown by \citet{dunham12} and \citet{vorobyov13}.  However, only a 
handful of synthetic observables from 3D magnetohydrodynamics (MHD) 
simulations have been presented in the literature 
\citep[e.g.,][]{commercon12a,commercon12b}.

Continuum observations probe the dust thermal emission and the dust 
structure around the protostar.  However, high spatial and spectral 
resolution molecular line observations are needed to probe the 
kinematical structure as the disk forms.  The aim of this paper is to 
present high-spatial (down to $0.1 \arcsec$ or 14 AU at a typical 
distance of 140 pc) synthetic observations of continuum and molecular 
lines from two of the 3D MHD collapse simulations presented in 
\citet{zyli13}.  The two simulations differ in the initial magnetic 
field direction with respect to the rotation axis: aligned and 
strongly misaligned where the magnetic field vector is perpendicular 
to the rotation axis.  The two cases represent the two extremes of 
the field orientation.  The synthetic observations will be compared 
with those from a 2D semi-analytical disk formation model presented in 
\citet{visser09} to investigate whether the predicted observables 
differ.  The 2D models allow us to simplify the different input 
parameters of the MHD simulations into two parameters: sound speed 
($c_{\rm s}$) and rotation rate.   Rotational transitions of CO are 
simulated to trace the observable kinematical signatures.

Another motivation in simulating CO molecular lines is the 
availability of high-quality spectrally and spatially resolved 
observational data toward embedded young stellar objects (YSOs) on 
larger scales ($> 1000$ AU).  Spectrally resolved lines have been 
obtained for low-excitation transitions $J_{\rm u} \leq 7$ ($E_{\rm 
u} = 155$ K) using ground-based facilities 
\citep[e.g.,][]{jorgensen02, vankempen09c, vankempen09d} and higher 
excited lines up to $J_{\rm u} = 16$ ($E_{\rm u} = 660$ K) using {\it 
Herschel}-HIFI \citep{hifi} in beams of 9--20$\arcsec$ 
\citep{yildiz10, yildiz13, kristensen13}.  Interestingly, 
\citet{sanjose-garcia13} found that the \ceo\ 9--8 lines are broader 
than the 3--2 lines for low-mass YSOs.  The observed line widths are 
larger than that expected from the thermal broadening, which indicates 
a significant contribution of microscopic turbulence or some other 
forms of motion, such as rotation and infall.  Through the 
characterization of spectrally resolved molecular lines on such 
physical scales, we aim to test the kinematics predicted in various 
star and disk formation models.

Another key test of star formation models is to compare the predicted 
mass evolution from the envelope to the star with observations 
\citep[e.g.,][]{prosac09, zyli14}.  Inferring these properties toward 
embedded YSOs is not straightforward due to the confusion between 
disk and envelope.  The mass evolution of the disk and envelope can 
be deduced from millimeter surveys combining both aperture synthesis 
and single dish observations \citep{km90, terebey93, looney03, 
prosac09, enoch11}.  However, precise determination of stellar masses 
requires spatially and spectrally resolved molecular line 
observations of the velocity gradient in the inner regions of 
embedded YSOs \citep{sargent87, ohashi97a, brinch07a, lommen08, 
prosac09, takakuwa12, yen13}.  Here we apply a similar analysis on 
the synthetic continuum and molecular line data as performed on the 
observations to test the reliability of the inferred masses.

This paper is structured as follows.  Section~\ref{sec:method} 
describes the simulations and the radiative transfer method that are
used.  The synthetic continuum images are presented in 
Section~\ref{sec:contmaps}.  Section~\ref{sec:colines} presents the 
synthetic CO moment maps and line profiles for the different 
simulations.  The results are then discussed in Section~\ref{sec:dis}
and summarized in Section~\ref{sec:sum}.


\section{Numerical simulations and radiative transfer}
\label{sec:method}

\begin{table*}[tbhp]
 \centering
 \caption{Stellar ($M_{\star}$), envelope ($M_{\rm env}$), and disk 
($M_{\rm d}$) masses for the three simulations.  $R_{\rm d}$ is the 
extent of the rotationally supported disk for each simulation. }
 \label{tbl:params}
 \begin{tabular}{l c c c c c c}\toprule\hline
 Model & $M_{\star}$ & $M_{\rm env}$ & $\Omega$ & $\lambda$ & $M_{\rm 
d}$ & $R_{\rm d}$ \\ 
  & [$M_{\odot}$] & [$M_{\odot}$] & [s$^{-1}$] & & [$M_{\odot}$] 
& [AU] \\ \hline
  3D MHD RSD    & 0.38 & 0.29 & $10^{-13}$ & 10 & 0.06 & 250--300 \\
  3D MHD No RSD & 0.24 & 0.35 & $10^{-13}$ & 10 & 
0.13\tablefootmark{*} & ...  \\
  2D RSD        & 0.35 & 0.32 & $10^{-13}$ & ... & 0.04 & 65  \\ 
\hline
  \end{tabular}
\tablefoot{\tablefoottext{*}{Mass of pseudo-disk is the sum of the 
regions with number densities $n_{\rm H2} > 10^{7.5}$ cm$^{-3}$; no 
radius is tabulated for this case.}}
\end{table*}

\subsection{Magneto-hydrodynamical simulations} \label{sec:mhd}

\begin{figure*}[htbp]
\centering
\includegraphics{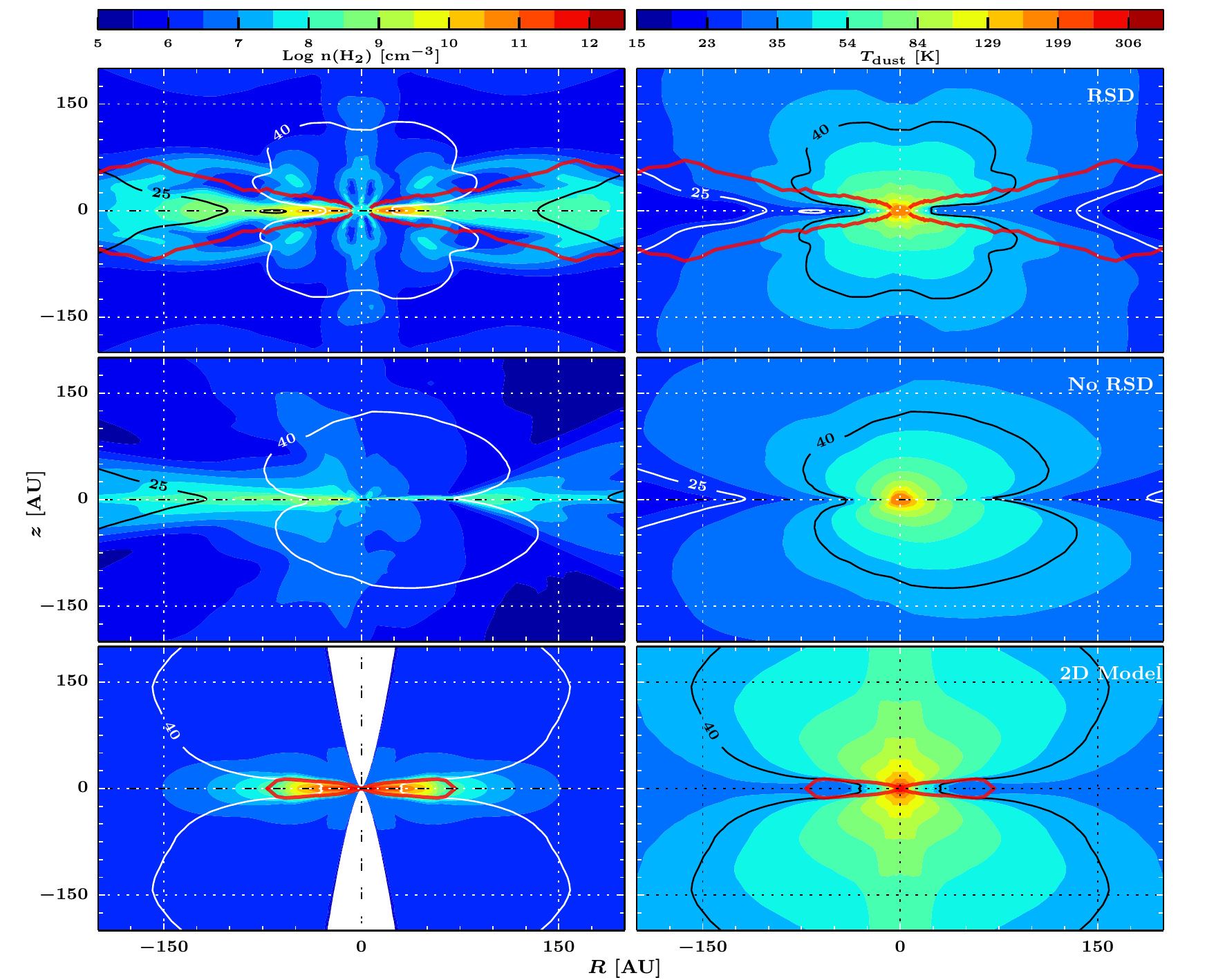}
\caption{Density and dust temperature structures in the inner 200 AU 
radius for all three simulations at $t\sim 1.2 \times 10^{5} \ {\rm 
years}$.  Temperature contours at 25 K and 40 K are indicated in the 
right panels.  {\it Top:}~A vertical slice ($R-z$ slice at $\phi=0$ 
where $R$ denotes the cylindrical radial coordinate) of the 3D MHD 
simulation of RSD formation.  {\it Middle:}~A vertical slice of the 3D 
MHD simulation of a pseudo-disk (No RSD).  {\it Bottom:}~2D 
semi-analytical disk formation model.  The red lines highlight the 
region of the stable RSD. }
\label{fig:phys2d}
\end{figure*}

\begin{figure*}[tbhp]
\centering
\includegraphics{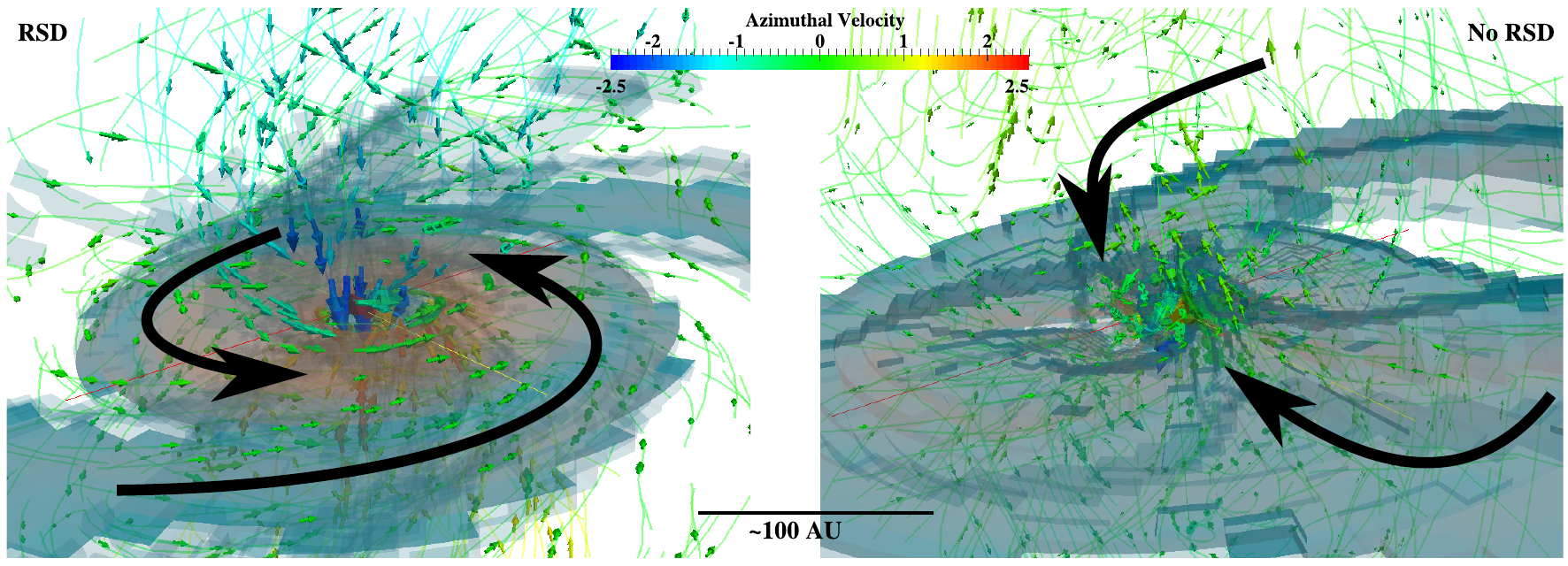}
\caption{Velocity streamlines for the two MHD simulations in 
\citet{zyli13}.  The simulations are rendered with \emph{paraview} 
(\url{http://www.paraview.org}).  {\it Left}: 3D MHD simulation of a 
collapsing uniform sphere with the magnetic field vector 
perpendicular to the rotation axis.  {\it Right}: The simulation with 
magnetic field vector aligned with the rotation axis.  The color of 
the arrows indicate the azimuthal velocity vector, $\upsilon_\phi$, 
in \kms.  The solid black arrows indicate the general stream lines in 
the two simulations.  The dark shaded colors indicate the density 
isocontours of $n_{\rm H_2} = 10^{7.5}$ cm$^{-3}$ up to $10^{10.5}$ 
cm$^{-3}$ (red). }
\label{fig:density}
\end{figure*}

We utilize the 3D MHD simulations of the collapse of a 1 $M_{\odot}$ 
uniform, spherical envelope as described in \citet{zyli13}.  The 
envelope initially has a density of $\rho_0 = 4.77 \times 10^{-19} {\ 
\rm g \ cm^{-3}}$, a solid-body rotation of $\Omega_0 = 10^{-13} \ 
{\rm Hz}$, and a relatively weak, uniform magnetic field of $B_0 = 11 
\ {\rm \mu G}$ ($\lambda$ = 10 where $\lambda$ is the dimensionless 
mass-to-flux ratio). The details of the simulations can be found in 
\citet{zyli13}.

A snapshot of two simulations at $t = 3.9 \times 10^{12}\ {\rm s} = 
1.24 \times 10^{5} \ {\rm years}$ is used.  This corresponds to near 
the end of the Stage 0 phase of star formation where almost one 
half of the initial core mass has collapsed onto the star 
\citep{robitaille06, dunham14}.  The difference between the two 
simulations is the tilt angle between the rotation axis and the 
direction of initial magnetic field vector, $\theta_0$.  One 
simulation starts with an initial tilt angle $\theta_0 = 0^{\circ}$ 
in which a pseudo-disk forms but not an RSD.  The other simulation 
starts with an initial tilt angle of $\theta_0 = 90^{\circ}$ in which 
an RSD forms \citep[see Fig.~\ref{fig:phys2d} and Figure 1 
in][]{zyli13}.

\begin{figure}[t]
 \centering
 \includegraphics{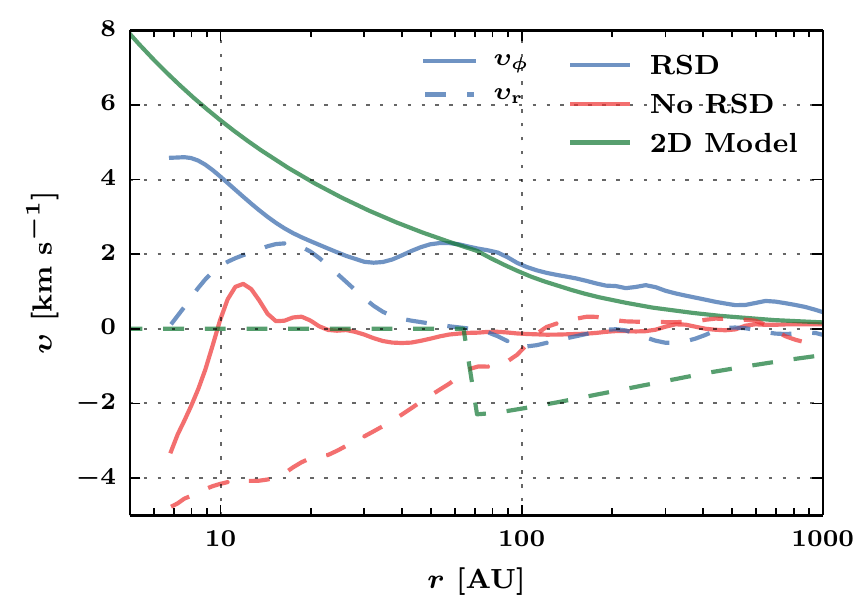}
 \caption{Radial ({\it solid}) and azimuthal velocities ({\it 
dashed}) at the midplane ($\theta = \pi/2$, $\phi = 0$) for the three 
different models: 3D RSD MHD simulation ({\it blue}), 3D No RSD ({\it 
red}), and 2D model ({\it green}). } 
\label{fig:simvelprofs}
\end{figure}

The RSD simulation ({\it left}) forms a flattened structure with 
number gas densities $n_{\rm H_{2}} > 10^{6.5}$ cm$^{-3}$ in the 
inner 300 AU radius.   In the region $r > 100$ AU, the magnitude of 
the radial and azimuthal velocities are within a factor of 2 of each 
other.  The radial velocities nearly vanish in the inner 70 AU radius 
(see Fig.~\ref{fig:simvelprofs}).  The streamlines in the RSD 
simulation show a coherent flattened rotating component (see 
Fig.~\ref{fig:density}).  In the case of the pseudo-disk simulation, 
number densities of $n_{\rm H_{2}} > 10^{6.5}$ cm$^{-3}$ encompass $r 
< 700$ AU regions, which is a factor of 2 larger than the RSD 
simulation.  An outflow cavity is present in this simulation with an 
expanding velocity field as shown in Fig.~1 ({\it Left}) of 
\citet{zyli13}.  The cavity is more evacuated compared with that in 
the simulation simulation that forms a rotationally supported disk 
although still not a canonical definition of a cavity.  The magnitude 
of the radial velocities are much larger than the azimuthal 
velocities in the inner $r < 300$ AU along most $\phi$ directions.  In 
this simulation, the streamlines show infalling material straight from 
the large-scale envelope onto the forming star.  Using these two 
simulations with very  different outcomes, we can investigate the 
similarities and differences in both continuum and molecular line 
profiles for pseudo-disk and RSD formation in 3D.


\subsection{Semi-analytical model} \label{sec:2dmodel}

For comparison, synthetic images from 2D semi-analytical axisymmetric 
models of collapsing rotating envelope and disk formation as 
described in \citet{visser09} with modifications introduced in 
\citet{visser10} and \citet{harsono13} are also simulated.  These 
models are based on the collapse and disk formation solutions of 
\citet{tsc84}, and \citet{cm81} including a prescription of an 
outflow cavity.  The disk evolution follows the $\alpha$-disk 
formalism as described in \citet{ss73} and \citet{lp74}.  The disk 
surface is defined by hydrostatic equilibrium as described in 
\citet{visser10} and is assumed to be in Keplerian rotation.  In order 
to compare with the MHD simulations, we consider the collapse of 1 
$M_{\odot}$, $c_{\rm s} = 0.26$ \kms, and $\Omega_0 = 10^{-13}$ Hz 
core.  The sound speed in this case is higher than that used in 
\citet{zyli13}, which affects the final disk properties at the end of 
the formation process.  The synthetic observables are produced at 
$t=3.9 \times 10^{12} \ {\rm s}$.  The bottom of 
Fig.~\ref{fig:phys2d} shows the physical structure of the 2D 
semi-analytical model at the time when a $\sim 65$ AU radius RSD is 
present.

A major difference between the 2D semi-analytical axisymmetric model 
and the 3D simulations is the outflow cavity.  The photon 
propagation is still treated in 3D.  Although outflowing gas is 
present in the pseudo-disk simulation (No RSD), the cavity remains 
filled with high number density ($10^5$ cm$^{-3}$) gas while lower 
density ($10^{2-3}$ cm$^{-3}$) gas occupies the cavity in the 2D 
model.  The outflowing gas generated from angular momentum conserving 
gas in the pseudo-disk model has relatively low velocities such that 
it does not clear out the cavity.  As a result, the dust temperature 
along the cavity wall is higher in the 2D model due to the direct 
illumination of the central star.  This is readily seen in the 40 K 
contour in Fig.~\ref{fig:phys2d} where it is elongated in the $z$ 
direction in the 2D case.  However, we show in Section 4 that 
this difference does not affect the results of this paper.


\subsection{Rotationally supported disk sizes and masses} 
\label{sec:rsd}

The extent of the RSD in the 2D semi-analytical model is  
defined by hydrostatic equilibrium.  Using these properties, the RSD 
is defined by a region with densities $> 10^{7.5}$ cm$^{-3}$ and 
azimuthal velocities $\upsilon_{\phi} > 1.2$ \kms.  The disk 
evolution in 2D follows the alpha-disk formalism with $\alpha = 
10^{-2}$, which in turns define the radial velocities as 
$\upsilon_{\rm r}/\upsilon_{\phi} \sim \alpha (H/R)^2 \sim 10^{-3}$ 
where $H$ is the disk's scale height.  Similar criteria are used to 
extract the extent of the RSD in the 3D simulation with the additional 
constraint of $v_{\phi} > v_{\rm r}$.  Note that we apply the criteria 
from the 2D model to define the RSD in the 3D simulation, thus it is 
is not necessarily in hydrostatic equilibrium.  This can be clearly 
seen in 
Fig.~\ref{fig:simvelprofs} for the 3D simulation where 
$\upsilon_{r}/\upsilon_{\theta}$ does not satisfy the hydrostatic 
disk criterion along a few azimuthal angles. Using these criteria, an 
RSD up to 260 AU is found in the mis-aligned simulation.  The extent 
of the 3D RSD is 300 AU if $v_{\phi} > 1$ \kms\ is used.  With the 
former criterion, the disk masses contained within such a region are 
0.06 $M_{\odot}$ for the 3D RSD and 0.04 $M_{\odot}$ for the 2D RSD 
surrounding 0.38 $M_{\odot}$ and 0.35 $M_{\odot}$ stars, respectively 
(see Table~\ref{tbl:params}).  As Fig.~\ref{fig:simvelprofs} shows, 
the radial velocity component of the pseudodisk is a significant 
fraction of the Keplerian velocity.  Due to such high radial 
velocities, the surface density of the pseudodisk remains low.  
However, the total mass of the high density regions ($n_{\rm H_2} > 
10^{7.5} \ {\rm cm^{-3}}$) is a factor of 2 higher than the RSD mass.


\subsection{Observables and radiative transfer}\label{sec:radtrans}

The first step before producing observables is the calculation of the 
dust temperature structure, which is critical for the molecular 
abundances since it controls the freeze-out from the gas onto the 
dust.  The dust temperature is computed using the 3D continuum 
radiative transfer code RADMC3D\footnote{\url{
http://www.ita.uni-heidelberg.de/\textasciitilde 
dullemond/software/radmc-3d}} with a central temperature of 5000 K, 
which is the typical central temperature in the 2D semi-analytical 
models at around the end of Stage 0 phase.  The central luminosity is 
fixed at 3.5 $L_{\odot}$ for all models.  The dust opacities used are 
those corresponding to a mix of silicates and graphite grains covered 
by ice mantles \citep{crapsi08}.  The 3D MHD simulation from 
\citet{zyli13} has an inner radius of 10$^{14}$ cm (6.7 AU) while the 
semi-analytical model has an inner radius of 0.1 AU.  The simulated 
observables presented here are not sensitive to the physical and 
chemical structures in the inner 10 AU radius.  Thus, these 
differences do not affect the conclusions of this paper.  The same 
opacities and central temperature are adopted for all simulations in 
order to focus on the general features of the observables.  The gas 
temperatures are assumed to be equal to the dust temperatures, which 
is valid for the optically thin lines simulated here that trace the 
bulk mass where $T_{\rm gas} \sim T_{\rm dust}$ \citep{doty02, 
doty04}.

\paragraph{CO abundance.}~In this paper, we concentrate on simulating 
CO molecular lines of $J =$ 2--1, 3--2, 6--5, and 9--8.  For 
simplicity, the \mco\ abundance is set to a constant value of 
$10^{-4}$ with respect to H$_2$ except in regions with $T_{\rm dust} 
< 25$ K, where it is reduced by a factor of 20 to mimic freeze-out 
\citep{jorgensen05a, yildiz13}.  We adopt constant isotopic ratios of 
$^{12}$C$/^{13}$C$=70$, $^{16}$O$/^{18}$O$=540$, and 
$^{18}$O$/^{17}$O$=3.6$ \citep{wilson94} to compute the abundance 
structures of the isotopologs.

\paragraph{Synthetic images.}~This paper presents synthetic continuum 
maps at 450, 850, 1100, and 1300 $\mu$m.  The images are rendered 
using RADMC3D with an image size of 8000 AU at scales of 5 AU pixels.  
They are placed at a distance of 140 pc.  Synthetic images at 
inclinations of $0^{\circ}$ (face-on; down the $z$-axis), 
$45^{\circ}$, and $90^{\circ}$ are produced.  The latter option is 
included because one of the claimed embedded disk sources is close to 
edge-on \citep[$i\sim 90^{\circ}$, L1527 in][]{tobin12}.  For the 
synthetic molecular lines, the local thermal equilibrium (LTE) 
population levels are computed using the partition functions adopted 
from the HITRAN database \citep{hitran}.  LTE is a good assumption 
because the densities in the simulations are greater than the 
critical densities of the simulated transitions.  Non-LTE effects may 
play a minor role for synthetic $J_{\rm u} \ge 6$ lines from the 
pseudo-disk simulation due to its lower densities relative to the 
other two simulations.  The line optical depth ($\tau_L$) is also not 
expected to play a role since the focus of this paper is on the minor 
isotopologs and on the kinematics dominated by the line wings where 
their line optical depths are lower than at the line center.  The 
properties of the molecules ($E_{\rm up}$ and $A_{\rm ul}$) are 
taken from the LAMDA database \citep{lamda}.  Only thermal broadening 
is included in simulating the molecular lines without any additional 
microturbulence.  The image cubes are rendered at a spectral 
resolution of 0.1 \kms\ covering velocities from -7.5 to 7.5 \kms.  
In order to simulate observations, the synthetic images are then 
convolved with Gaussian beams between 0.1$\arcsec$ to 20$\arcsec$.  
The convolution is performed in the Fourier space with normalized 
Gaussian images.


\section{Continuum} \label{sec:contmaps}

\subsection{Images and prospects for ALMA}

\begin{figure}[htbp]
\centering
\includegraphics{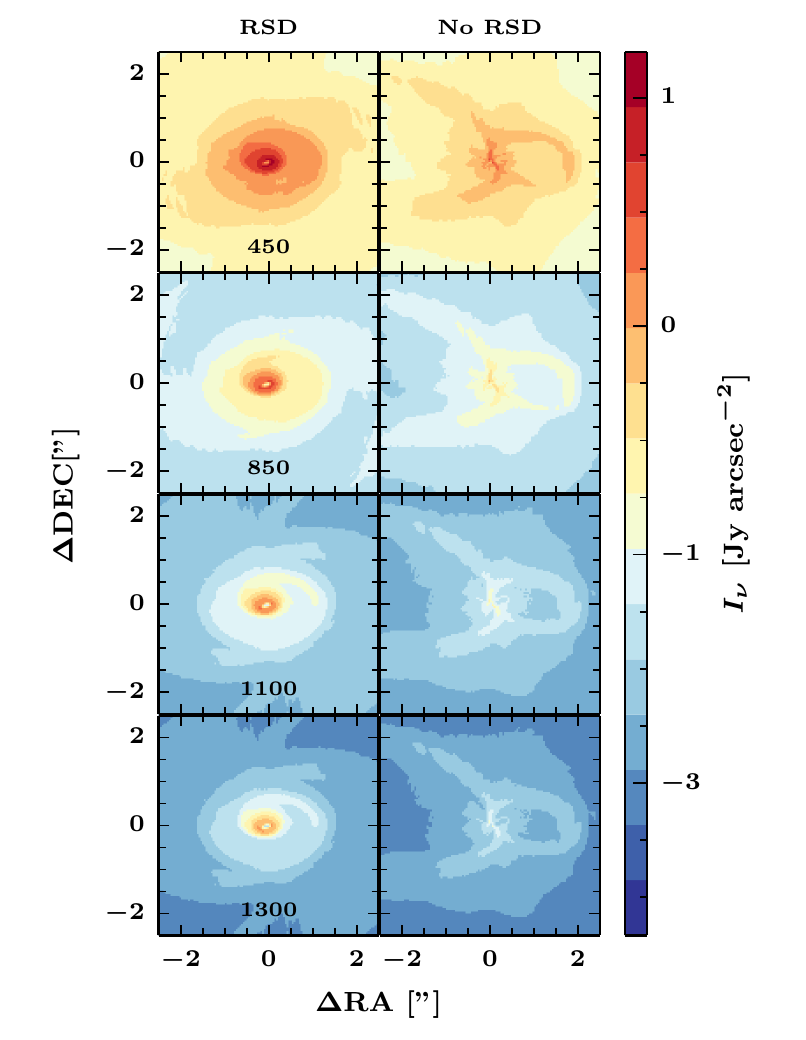}
\caption{Continuum intensity maps of 450, 850, 1100, and 1300 $\mu$m 
for the MHD disk formation simulation (left panels) and pseudo-disk 
(No RSD, right panels) at $i = 45^{\circ}$ at 5 AU pixels.  Note the 
large dynamic range needed to see all of the structures.}
\label{fig:unconvcont}
\end{figure}

\begin{figure*}[htbp]
\centering
\begin{tabular}{c c}
\includegraphics{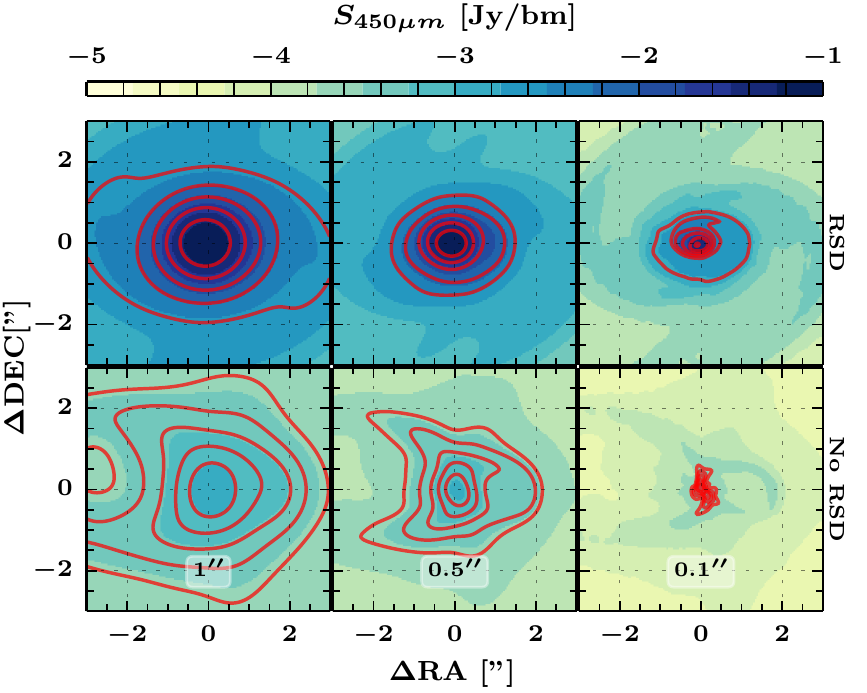}
&
\includegraphics{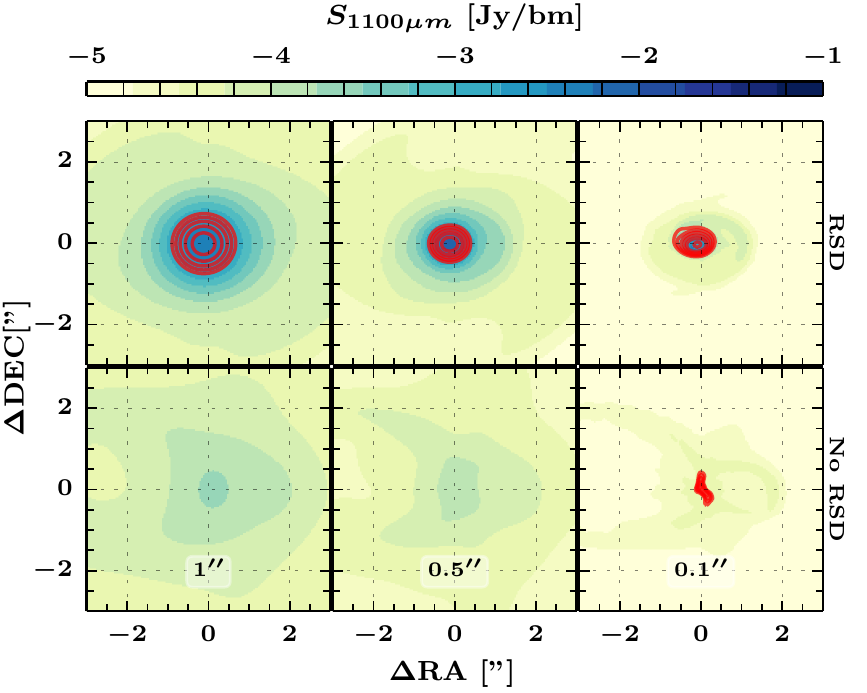}
\end{tabular}
\caption{Convolved continuum maps in the inner 5$\arcsec$ of 450 
({\it left}) and 1100 ({\it right}) $\mu$m at $i = 45^{\circ}$.  The 
images are convolved with $1\arcsec$, $0.5\arcsec$, and $0.1\arcsec$ 
beams as indicated in each panel.  For each panel, the top row shows 
synthetic image of RSD simulation and the pseudo-disk simulation is 
shown in the bottom row.  The color scale presents the full range of 
the emission above $10^{-2}$ mJy/bm, not all of which may be 
detectable.  Solid red contours are drawn from $3\sigma$ up to maximum 
at 6 logarithmic steps where $1\sigma$ is 0.01 $\times$ the maximum 
(dynamical range of 100) with a minimum at 0.5 mJy/bm for the 
$1\arcsec$ and $0.5\arcsec$ images.  For the images convolved with a 
$0.1\arcsec$ beam, the red contours are drawn with a minimum noise 
level of $0.5$ mJy/bm for 450 $\mu$m and $0.05$ mJy/bm at longer 
wavelengths with a dynamic range of 1000, as appropriate for the full 
ALMA.}
\label{fig:convimage}
\end{figure*}

Continuum images are rendered at four wavelengths and viewed at three 
different inclinations.  Figure~\ref{fig:unconvcont} presents the 
synthetic 450, 850, 1100 and 1300 $\mu$m continuum images at an 
inclination of 45$^{\circ}$ for the two 3D MHD simulations.  The left 
panels present the images from 3D RSD simulation while the right 
panels show images from the pseudo-disk simulation (labeled as No 
RSD).  The features produced during the collapse are clearly visible 
in the 450 $\mu$m map; most of them are two orders of magnitude 
fainter at 1300 $\mu$m.  The spiral structure in the 450 $\mu$m image 
is due to magnetically channelled, supersonically collapsing material 
on its way to the RSD, rather than a feature of the RSD itself.

One of the aims of this paper is to investigate whether these 
features are observable with current observational limits and what is 
possible with future Atacama Large Millimeter/submillimeter Array 
(ALMA) data. With the full ALMA, a sensitivity of $\sigma$=0.5 mJy 
bm$^{-1}$ (bm = beam) at $\sim 450$ $\mu$m and 0.05 mJy bm$^{-1}$ at 
1100 $\mu$m can be achieved at spatial resolutions $\leq 0.1\arcsec$ 
for 30 minutes of integration (1.8 GHz bandwidth).  In addition, a 
high dynamic range of $> 1000$ ($\sigma = \ 0.001\times S_{\rm peak}$) 
can also be achieved.  Figure~\ref{fig:convimage} presents the images 
convolved with a $0.1\arcsec$ beam for the two 3D MHD simulations 
(right-most panels).  The color scale indicates the full range of 
emission, while the red solid lines show the region above 3$\sigma$ 
where $\sigma$ is either dynamically limited to 1000 or to the 
$\sigma$ values listed above.  This simply means that the solid red 
lines indicate the detectable features.  With a combination of high 
dynamic range and sensitivity, the features of the collapse are 
observable and distinguishable in both 450 $\mu$m and 1100 $\mu$m 
continuum maps.  At a spatial resolution of $0.1 \arcsec$, most of 
the emission at 1100 $\mu$m is due to the rotationally supported disk 
with a small contribution from the surrounding envelope.

As Figs.~\ref{fig:unconvcont} and ~\ref{fig:convimage} show, both the 
strength of the features and their extent change with wavelength, as 
indicated by the red line contours.  With a resolution of 
$0.1\arcsec$, the extent of the detectable emission decreases from 
$1.5\arcsec$ at 450 $\mu$m to $<1\arcsec$ at 1100 $\mu$m for the RSD 
case.  At long wavelengths, the dust emission is given by $I_{\nu} 
\sim T_{\rm dust} \nu^2 \times (1 - e^{-\tau_{\rm dust}})$.  Since the 
dust emission is optically thin at long wavelengths, the intensity is 
$\propto T_{\rm dust} \nu^2 \tau_{\rm dust}$.  The variables that 
depend on position are $T_{\rm dust}$ and $\tau_{\rm dust} \propto 
\rho_{\rm dust} \kappa_{\nu}$.  The frequency dependence of the 
opacity is $\kappa_{\nu} \propto \nu^{1.5}$ for the adopted opacity 
table.  The extent of the detectable emission depends on these 
quantities.  At one particular position, the ratio of the emission at 
the two wavelengths is simply $I_{\rm 450 \mu m} / I_{\rm 1100 \mu m} 
\propto \nu^{3.5}$.  Thus, the predicted difference in size at the 
two wavelengths is due to the frequency dependence of the emission.

For ALMA early science observations (cycles 0 and I), the 
capabilities provided a dynamic range only up to 100 at a spatial 
resolution of $\sim 0.5\arcsec$.  Figure~\ref{fig:convimage} presents 
synthetic 450 and 1100 $\mu$m images convolved with 1$\arcsec$ and 
0.5$\arcsec$ beams, compared with the 0.1$\arcsec$ beam images.  
The color scale again indicates the full range of emission, while the 
red solid lines now show the region above 3$\sigma$ where the noise 
level, $\sigma$, is dynamically limited to 100 with a minimum of 
$0.5$ mJy/beam at both wavelengths.  The red lines again present the 
observable emission.  The 450 $\mu$m images convolved with a 
$1\arcsec$ beam show an elongated flattened structure for both RSD 
and pseudo-disk simulations and are therefore indistinguishable.  
Most of the emission at 1100 $\mu$m and longer wavelengths is not 
detectable at the assumed noise level of $0.5$ mJy bm$^{-1}$.  A 
similar result is found after convolution with a $0.5\arcsec$ beam.  
Although the synthetic observations from the pseudodisk indicate a 
`cometary' structure, this may be affected by the presence of outflow 
cavity which is absent in the case of RSD.  Thus, the full ALMA 
capabilities are needed to distinguish models based on continuum data 
only.


\subsection{Inclination effects} \label{sec:inclcont}

\begin{figure}[htbp]
\centering
\includegraphics{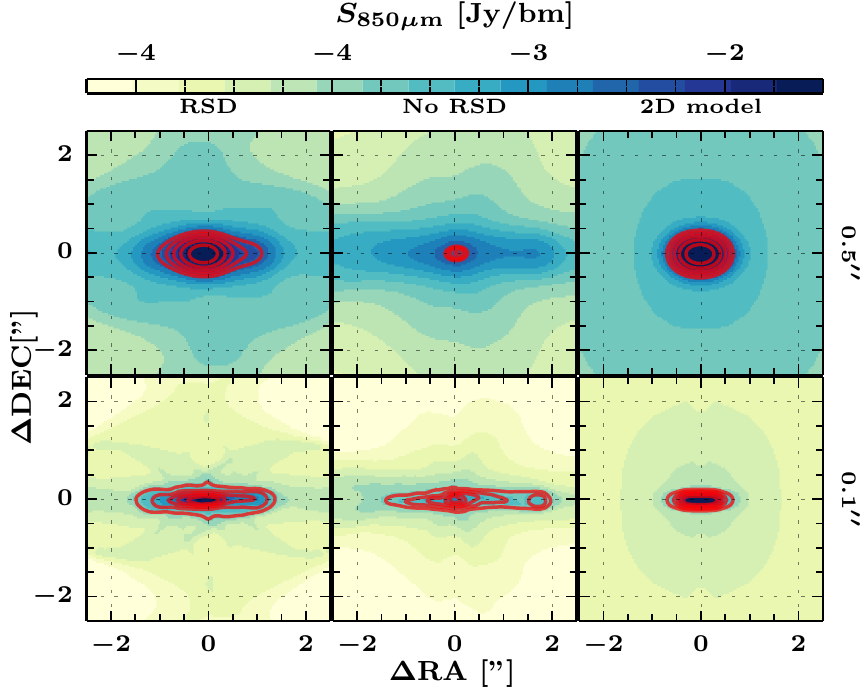}
\caption{Synthetic 850 $\mu$m continuum images convolved with 
$0.5\arcsec$ ({\it top}) and $0.1\arcsec$ ({\it bottom}) beams for 
the three simulations viewed at $90^{\circ}$ (edge-on): RSD formation 
({\it left}), pseudo-disk ({\it center}), and 2D semi-analytical model 
({\it right}).  The red solid lines are drawn at the same contours as 
in Fig.~\ref{fig:convimage}. }
\label{fig:modelat90}
\end{figure}

The images for a face-on system ($i=0^{\circ}$) are similar to those 
of the moderately inclined system ($i = 45^{\circ}$) presented in 
Fig.~\ref{fig:convimage}.  For low inclinations between $0$ to 
$45^{\circ}$, the continuum images only change slightly in terms of 
absolute flux density and show a small elongation due to orientation.

In contrast, the two simulations rendered at edge-on ($i \sim 
90^{\circ}$) geometry exhibit similar compact components even 
at the highest angular resolution (Fig.~\ref{fig:modelat90}).  They 
both show an elongated flattened disk-like emission similar to the 2D 
semi-analytical model.  This signature suggests an RSD, but it is due 
to the pseudo-disk produced in the 3D MHD simulation whose initial 
magnetic field vector is aligned with rotation axis.  The peak 
continuum emission of the pseudo-disk is a factor of 10 lower than 
that of the other two models while it is similar between the 3D RSD 
and the 2D semi-analytical models (difference of $<10\%$).  The main 
difference is the extent of the elongated emission where the 2D model 
predicts a very compact ($\sim 1 \arcsec$ radius) structure while the 
pseudo-disk component shows an extended flattened structure ($\le 2 
\arcsec$).  This illustrates the difficulties in testing disk 
formation models for highly inclined systems based solely on 
continuum data.


\section{Molecular lines} \label{sec:colines}

The continuum emission arises from thermal dust emission and does not 
contain kinematical information.  As shown in the previous section, a 
compact flattened structure is expected in the continuum maps in the 
inner regions of all three models.  Kinematical information as 
contained in spectrally resolved molecular lines is essential to 
distinguish the models and to derive stellar masses.  The rotational 
lines of CO isotopologs are used to investigate this.

The rotational lines of \tco, \ceo, and \cso\ are simulated.  CO is 
chosen since its abundance is less affected by chemical evolution 
during disk formation.  We do not investigate \mco\ lines since they 
are dominated by the entrained outflow material and are optically 
thick.  The isotopolog lines are more optically thin and are expected 
to probe the higher density region where the disk is forming.  These 
predicted spatially resolved molecular line maps can be compared with 
ALMA data.  Moreover, high quality spectrally resolved CO isotopolog 
lines probing the larger-scale envelope toward low-mass embedded YSOs 
have been obtained with single-dish telescopes (see \S 1).  The 
characterization of these \ceo\ and \cso\ line profiles provide a 
test for the kinematical and density structures of the collapsing 
protostellar envelope on larger scales \citep[e.g.,][]{hogerheijde98, 
jorgensen02}.


\subsection{Moment maps: RSDs or not?} \label{sec:mommap}

\begin{figure}[htbp]
\centering
\includegraphics{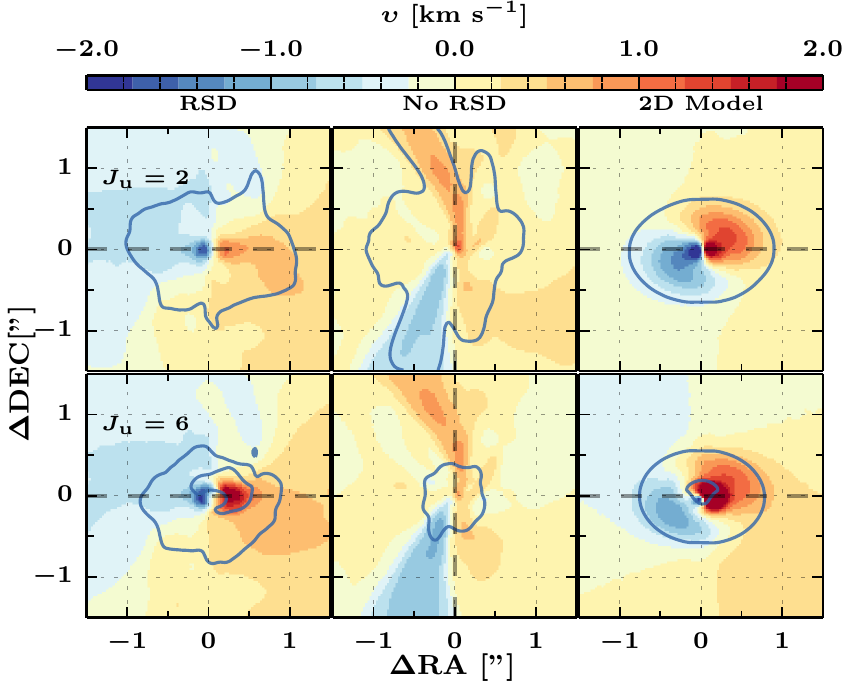}
\caption{Moment one maps of \ceo\ 2--1 ({\it top}) and 6--5 ({\it 
bottom}) for the three simulations viewed $i=45^{\circ}$ convolved 
with a $0.1\arcsec$ beam.  The solid line shows the 20\% intensity 
contour of the moment zero ($\int S_{\upsilon} d\upsilon$) peak map to 
indicate the flattened structure.  The black dashed lines indicate 
direction at which the PV slices are constructed and the direction of 
the elongation in the moment zero maps. }
\label{fig:moment13265}
\end{figure}

Observationally, the kinematical information of the infalling 
envelope and RSD is inferred through moment maps.  Elongated moment 
zero (velocity integrated intensity) maps give an indication of the 
presence of a flattened structure that is associated with a disk.  
Meanwhile, coherent velocity gradients in the moment one (velocity 
weighted intensity) map may point to a rotating component.  Analysis 
of synthetic moment maps of the simulations are presented and 
compared in this section.  We focus on presenting the synthetic 
`interferometric' maps of optically thin \ceo\ and \cso\ lines by 
convolving the image cubes with a $0.1\arcsec$ Gaussian beam.  The 
construction of moment maps only takes into account emission $> 
1$\% of the peak emission.  This translates to a noise level of 
$0.3$\% of the peak emission.  Although ALMA can achieve a dynamic 
range of $>500$, it is more likely that the molecular lines of minor 
isotopologs from low-mass YSOs will be noise limited at the line 
wings since they are weaker than the emission near the line center 
for typical observations with 1--2 hours integration time.

Figure~\ref{fig:moment13265} presents synthetic first moment 
(flux-weighted velocity) maps.  The moment zero contours at 20\% of 
peak intensity indicate the elongation direction.  The flattened 
density structure is oriented in the east-west direction (horizontal) 
for all three models as seen in Fig.~\ref{fig:phys2d}.  Coherent 
velocity gradients from blue- to red-shifted velocity are seen in all 
three models but not necessarily in the same direction.  The presence 
of an embedded RSD is revealed in the 3D RSD and 2D model by the 
coherent blue- to red-shifted velocity gradient in the east-west 
direction similar to the flattened disk structure.  On the other hand, 
in the pseudo-disk simulation, the velocity gradient is in the 
north-south direction similar to the continuum image as shown in 
Fig.~\ref{fig:convimage} (right-most panels).  Such velocity gradients 
can be mistaken to be along the major axis of the disk without higher 
spatial resolution and sensitivity data.

\begin{figure*}[tbhp]
\centering
\begin{tabular}{cc}
\includegraphics{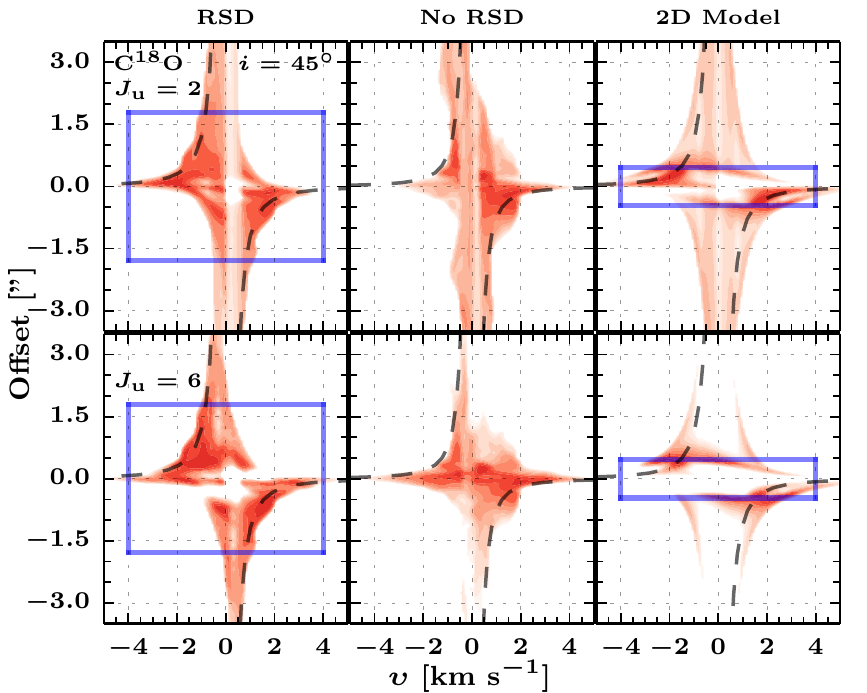}
&
\includegraphics{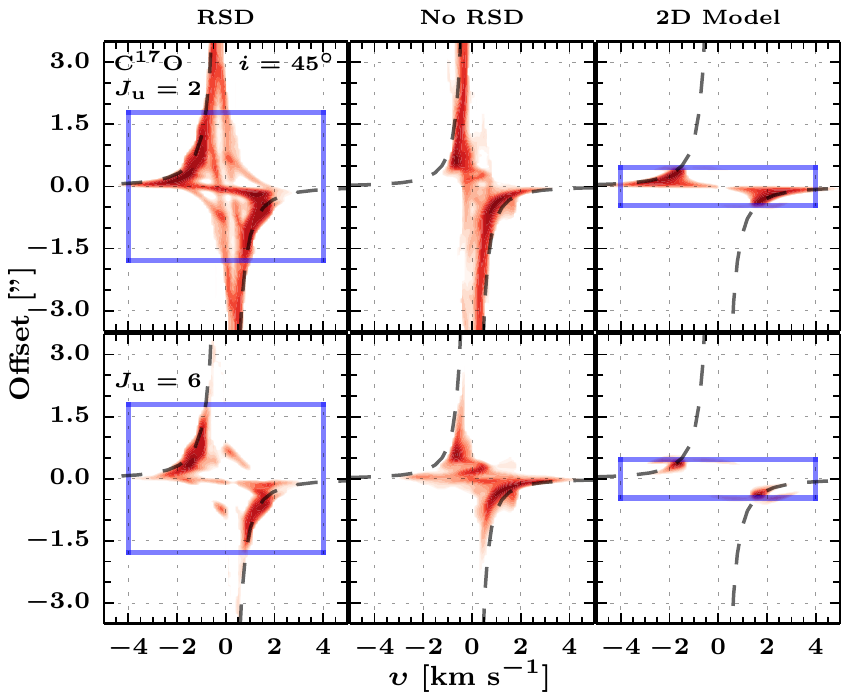}
\end{tabular}
\caption{PV maps of \ceo\ ({\it left}) and \cso\ ({\it right}) along 
the velocity gradients as seen in Fig.~\ref{fig:moment13265} for an 
inclination of $45^{\circ}$ and $0.1\arcsec$ beam.  The top panels 
show the $J=$ 2--1 line and the bottom panels show the 6--5 
transition.  The blue rectangle indicates the radii of RSDs.  The 
dashed lines indicate the inclination corrected Keplerian curves 
associated to the stellar mass.  The red color scales show emission 
from 10\% to the peak intensity.}
\label{fig:pvmap1}
\end{figure*}

A number of effects conspire to generate a velocity gradient along 
the minor axis of the flattened structure in the pseudo-disk 
simulation.  First, since the magnetic braking is efficient in this 
case, the dominant motion of the material is in the radial direction 
(see Figs.~\ref{fig:density} and \ref{fig:simvelprofs}).  Second, the 
flattened structure in this particular simulation has lower-density 
gas than the RSD simulation (see Section~\ref{sec:mhd}).  The $n_{\rm 
H_2} > 10^{6.5}$ cm$^{-3}$ region extends up to 700 AU in size and at 
an angle with respect to the rotation axis (see 
Fig.~\ref{fig:phys2d}).  Thus, the north-south direction in the 
pseudo-disk model shows the infalling material along the streamlines 
connecting the large-scale envelope and the central star.

At moderate inclinations, the pseudo-disk simulation therefore shows 
a coherent velocity gradient in a more-or-less straight north-south 
line.  The velocity gradient changes to an east-west direction at 
high inclinations.  At high inclinations, the observer has a direct 
line of sight on the high density region shown in 
Fig.~\ref{fig:phys2d} and therefore the line emissions pick up the 
rotational motions of the flattened structure similar to the RSD 
simulations.  Furthermore, the skewness that is present in the moment 
one maps of RSD simulations largely disappears at high inclinations 
for the same reason.  Thus, it is difficult to separate the envelope 
from the disk for high inclinations (almost edge on) from moment one 
map alone.


\subsection{Velocity profiles}\label{sec:velprof}

\subsubsection{PV cuts}

Observationally, the presence of embedded Keplerian disks is often 
established by constructing position velocity (PV) diagrams 
along the major axis of the system as seen in moment zero maps.  In 
theory, the PV analysis is straight-forward.  It is symmetric in both 
position and velocity space (4 quadrants are occupied) if the system 
is infall dominated \citep[e.g.,][]{ohashi97a, brinch08}.  The 
symmetry is broken if rotation is present and the emission peaks are 
shifted to larger offsets corresponding to the strength of the 
rotational velocities (2 quadrants are occupied).

Figure~\ref{fig:pvmap1} presents synthetic PV diagrams along 
the major axis of the disk where it corresponds to the direction of 
the blue- to red-shifted velocity gradient.  For the images in 
Fig.~\ref{fig:moment13265}, an east-west slice (horizontal) is taken 
for the RSD simulations, while a north-south (vertical) slice is 
adopted for the case of the simulation without an RSD (No RSD).  
These slices are not exactly the major axis of the moment zero map, 
however these directions pick up most of the velocity gradient 
present in the inner $1\arcsec$.  Both \ceo\ and \cso\ lines are 
simulated. Interestingly, the PV slices suggest that rotational 
motions are present regardless of whether an RSD is present or not.  
This is most readily seen in the \cso\ PV maps (right of 
Fig.~\ref{fig:pvmap1}) in which only 2 of the 4 quadrants are 
occupied by molecular emissions for all three models at $i = 
45^{\circ}$.  This shows that \cso\ emission readily picks up the 
rotational motion at small-scales but also that infalling motion can 
be confused with rotation if the wrong direction for the PV cut is 
chosen (No RSD model).

\begin{figure*}[tbhp]
\centering
\sidecaption
\begin{tabular}{c}
\includegraphics{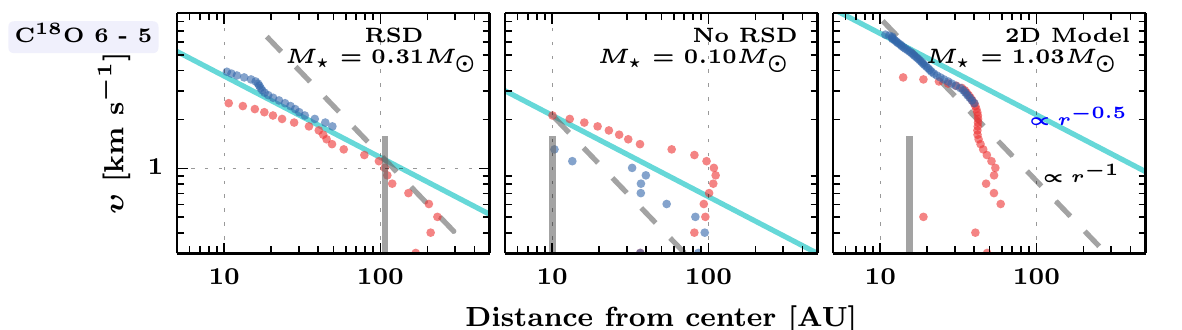}
\\
\includegraphics{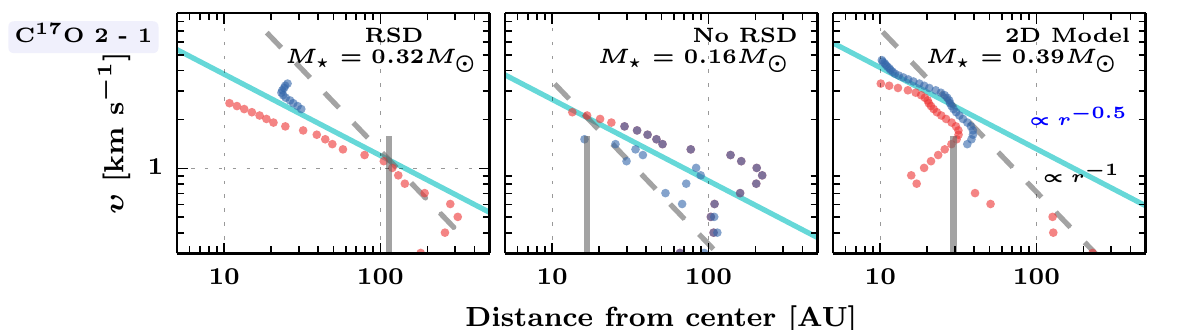} 
\end{tabular}
\caption{Peak position-velocity diagrams of \ceo\ 6--5 ({\it top}) 
and \cso\ 2--1 ({\it bottom}) at $i = $ 45$^{\circ}$ after 
convolution with a $0.1\arcsec$ beam.  The red and blue symbols 
correspond to the red- and blue-shifted velocity components,  
respectively.  The different panels present the PPV for the different 
simulations.  Vertical solid lines show the disk radii extracted for 
the three models and dashed lines show the steep velocity profile 
($\upsilon \propto r^{-1}$) while the solid cyan lines indicate the 
Keplerian curves.  The stellar masses are indicated in the top right 
of each panel.  The offset between blue- and red-shifted points are 
due to the limited spatial and spectral resolution at the high 
velocities.}
\label{fig:velprof1}
\end{figure*}

The \ceo\ PV maps indicate contributions from the infalling envelope 
since the maps are more symmetric than those in \cso.  The \ceo\ PV 
slices of the pseudo-disk simulations indicate an infall-dominated 
structure in which the 4 quadrants are occupied.  On the other hand, 
there is a clear indication of a rotating component for the two RSD 
simulations in which only 2 of the 4 quadrants are filled at small 
radii.  This suggests that spatially and spectrally resolved \ceo\ 
lines can distinguish between a pseudo-disk and an RSD.

Figures~\ref{fig:pvmap1} compares the PV maps of the 2--1 and 6--5 
transitions.  Most of the emission in the 6--5 transition occupies 
only 2 of the 4 quadrants indicating signatures of rotational 
motions, whereas the 2--1 lines also show some emission in the other 
2 quadrants from larger scales.  This is a clear indication that the 
6--5 line is a better probe of the rotational motions in the inner 
100 AU than the 2--1 line.  Yet, a combination of spatially and 
spectrally resolved molecular lines are needed to confirm the 
presence of embedded rotationally supported disk.

\subsubsection{Peak position-velocity diagrams}

While it is clear that there is indeed a rotating component for some 
models, the question is whether the extent of the Keplerian structure 
can be extracted from such an analysis.  A Keplerian rotating 
flattened structure exhibits a velocity profile $\upsilon\propto 
r^{-0.5}$, where $r$ is the distance from the central source.  These 
positions are either determined from fitting interferometric 
visibilities of each velocity channel \citep{lommen08, prosac09} or 
determination of the peak positions in the image space 
\citep{tobin12, yen13}.  We here determine the peak positions 
directly in the image space to assess whether a velocity profile is 
visible in the synthetic molecular lines.

Peak positions are determined for each of the velocity channel maps 
for each molecular line for the red and blue-shifted components 
separately taking into account channels whose peak flux density 
($S_{\nu}$ in Jy bm$^{-1}$) are $>1\%$ of $S_{\rm max}$.  They 
are subsequently rotated according to the direction of the velocity 
gradient.  If an RSD is present, the peak positions of both red- and 
blue-shifted velocities are expected to follow the Keplerian velocity 
profile ($\upsilon \propto r^{-0.5}$).  The combination of infalling 
rotating envelope and RSD, which exhibits a skewness in the moment 
one map, is expected to show a steeper velocity profile ($\upsilon 
\propto r^{-1}$) \citep{yen13}.  However, at high velocities, the 
peak positions of the red- and blue-shifted velocities can be 
misaligned at scales of 5 AU due to the limited spatial resolution.  
The disk radius is determined by minimizing the difference ($\sim 
10\%$) between the best-fit stellar mass inside and at the disk 
radius.

The peak position-velocity diagrams (PPVs) for \ceo\ 6--5 and \cso\ 
2--1 are shown in Fig.~\ref{fig:velprof1}. These lines are chosen 
because they represent two observational extremes in terms of 
excitation and optical depth.  There is a clear distinction between 
the RSD simulations and a pseudo-disk.  The velocity profile of the 
pseudo-disk is much flatter than that expected from an RSD.  Thus, 
spatially and spectrally resolved molecular lines observations can 
clearly differentiate between an RSD and a pseudo-disk.

All three PPVs indicate a velocity profile close to $\upsilon 
\propto r^{-0.5}$ (for the pseudo-disk see \cso\ 2--1 PPV in the 
inner 20 AU).  This reflects the fact that the inner 300 AU of the 
models is dominated by velocity structure that is proportional to 
$r^{-0.5}$, which is both the radial velocity $\left ( \upsilon_{\rm 
infall} \propto \sqrt{\frac{2 G M_{\star}}{r}} \right )$ and angular 
velocity $\left ( \upsilon_{\rm rot} \propto \sqrt{\frac{G 
M_{\star}}{r}} \right )$ \citep{brinch08}.  From such 
characterization, the stellar masses can be calculated and indicated 
in the top right corner of Fig.~\ref{fig:velprof1}.  In general, the 
best-fit stellar masses in the case of RSD are within 30\% of the true 
stellar masses tabulated in Table~\ref{tbl:params}.  This is not so 
for the 2D model in the \ceo\ 6--5 and also \cso\ 6--5 (not shown) 
due to the fact that the inner flattened envelope is warm ($> 40$ 
K) and dense.

Another parameter that one would like to extract is the disk radius, 
$R_{\rm d}$, as indicated by the vertical solid line in 
Fig.~\ref{fig:velprof1}.  The break at $R_{\rm d}$ is readily seen in 
the 2D model at $\sim 40$ AU for both \ceo\ 6--5 and \cso\ 2--1.  For 
the case of the 3D MHD simulation (RSD), the best-fit radius varies 
between 100 and 300 AU.  It also exhibits a steep velocity profile 
($\upsilon \propto r^{\sim -1}$) at radii $ > 100$ AU.  The large 
range of disk radii is due to the envelope emission overwhelms the 
molecular emission because CO is frozen out in the cold part of the 
disk at those large radii.  The issue of disk versus envelope 
emission becomes apparent in the case of large embedded disk ($R_{\rm 
d} > 100$ AU).

The comparison shows that there is a clear distinct PPV profile 
associated with RSD formation from $r^{-0.5}$ to $r^{-1}$.   A steep 
velocity profile ($\upsilon \propto r^{-1}$) is absent in the 
pseudo-disk simulation.  This seems to indicate that such a steep 
velocity profile describes on-going RSD formation based on the given 
simulations.  A pseudo-disk is characterized by a flat velocity 
profile in the inner regions.  Furthermore, the PPV method can 
simultaneously derive the stellar mass and the extent of the RSD 
while separating the infalling rotating envelope from it.  With 
respect to differentiating between RSD and non-RSD, PPV is a better 
tool than PV-diagrams.


\subsection{Single-dish line profiles}

\begin{figure}[htbp]
\centering
\includegraphics{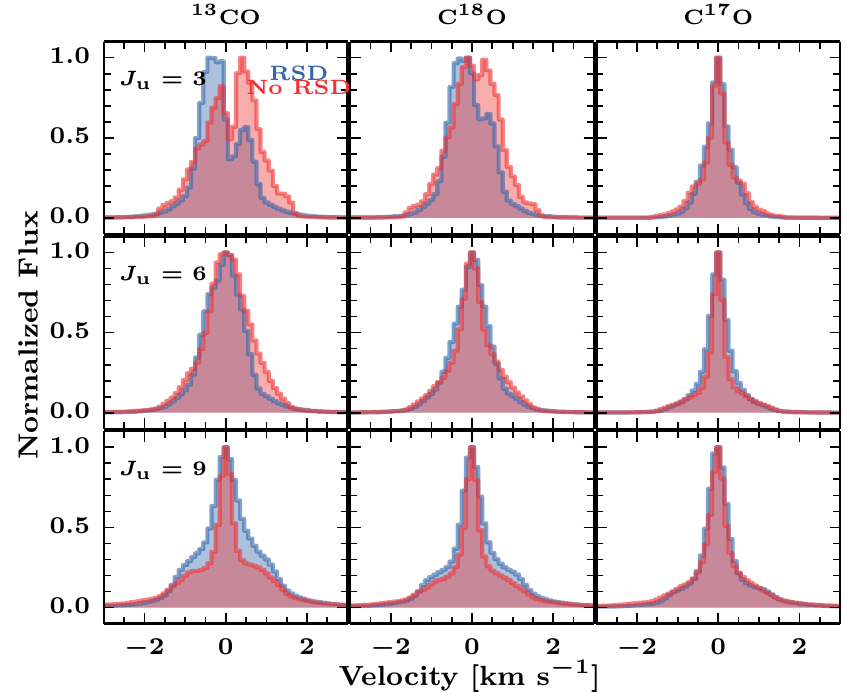}
\includegraphics{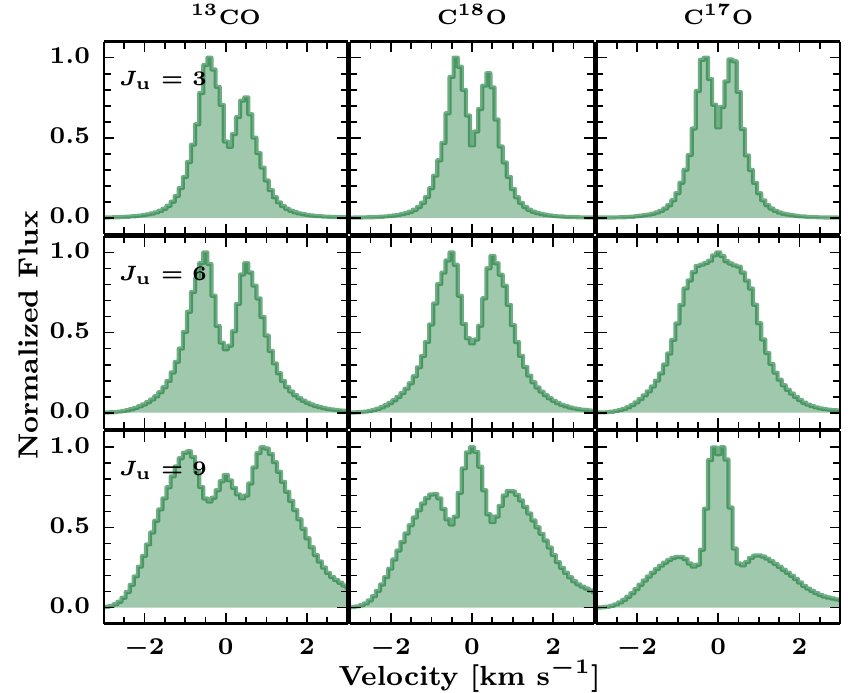}
\caption{ {\it Top: } \tco, \ceo, and \cso\ 3--2 ({\it top}), 6--5 
({\it middle}), and 9--8 ({\it bottom}) spectra at $i \sim 0^\circ$ 
within a 9$\arcsec$ beam.  The blue line shows the synthetic line from 
simulation with an RSD while the red line is the simulation without an 
RSD.  {\it Bottom: } Spectral lines convolved with a $9\arcsec$ beam 
simulated from 2D semi-analytical model viewed at face-on 
orientation.}
\label{fig:alllines9}
\end{figure}

The previous sections focus on features at small-scales as expected 
from interferometric observations.  The next assessment is to compare 
the synthetic molecular lines with single-dish observations which 
probe the physical structure of the large-scale envelope on scales up 
to a few thousand AU.  The image cubes are convolved with 3 different 
beams: 9$\arcsec$, 15$\arcsec$, and 20$\arcsec$.  These different 
beams are typical for single-dish CO observations using the  JCMT 
(15$\arcsec$), Atacama Pathfinder EXperiment (APEX, 9$\arcsec$), and 
{\it Herschel} (20$\arcsec$).  Figure~\ref{fig:alllines9} presents 
the synthetic CO lines ($J_{\rm u}$ = 3, 6, and 9) for the two MHD 
simulations viewed face-on ($i \sim 0^{\circ}$) convolved with a 
$9\arcsec$ beam.  The face-on orientation is considered first to 
compare with the line profiles in \citet{harsono13} for the 2D 
simulation.  The low-lying transitions ($J_{\rm u}$ = 3) probe the 
kinematics in the large-scale envelope.

Double peaked line profiles are present in the \tco\ and \ceo\ 3--2 
regardless whether an RSD is present or not.  For the \tco\ line, an 
inverse P-Cygni line profile is seen due to the coherent infalling 
material onto the disk while a P-Cygni profile is associated to the 
pseudo-disk, which is tracing the expanding material due to outflowing 
material present in the pseudo-disk simulation.  Self-absorption 
causes the double peak in the \tco\ 3--2 line due to optical depth 
(typically, $\tau_{\rm L} > 5$) at line center whereas it is weakly 
affecting the \ceo\ 3--2 line.

It is interesting to note that there is no significant difference in 
the 6--5 lines between a simulation that forms an RSD versus a 
pseudo-disk.  This transition ($E_{\rm u} \sim$110 K) traces the 
dense warm gas where a large fraction of the emission comes from $\ge 
40$ K gas \citep{yildiz10}.  The P-Cygni line profile is still 
visible in the \tco\ line, however it is not significant in the \ceo\ 
line.  The lines are also not Gaussian with significant wing emission 
extending up to $\pm 2$ \kms.

The line profiles for the semi-analytical models within a 9$\arcsec$ 
beam are shown in the bottom of Fig.~\ref{fig:alllines9}.  They are 
significantly different from the 3D MHD models, which arises from the 
prescribed velocity structure.  In \citet{harsono13}, an additional 
microturbulent broadening of $0.8$ \kms\ was added, which results in 
Gaussian line profiles consistent with the observed single-dish CO 
line profiles.  However, in this paper, we have not included 
the additional broadening term in order to investigate the emission 
arising from the true kinematical information.  The peaks are more 
prominent than those in the 3D MHD simulations due to a jump between 
the velocity structures of the RSD component and the infalling 
envelope.  In general, the 2D semi-analytical models produce 
significantly broader lines and significant variations between the CO 
isotopologs and transitions compared with 3D simulations because of a 
warmer disk and outflow cavity wall (see Section~\ref{sec:2dmodel}) 
which allow for stronger wing emissions.


\subsubsection{Inclination effects}

\begin{figure}[htbp]
\centering
\includegraphics{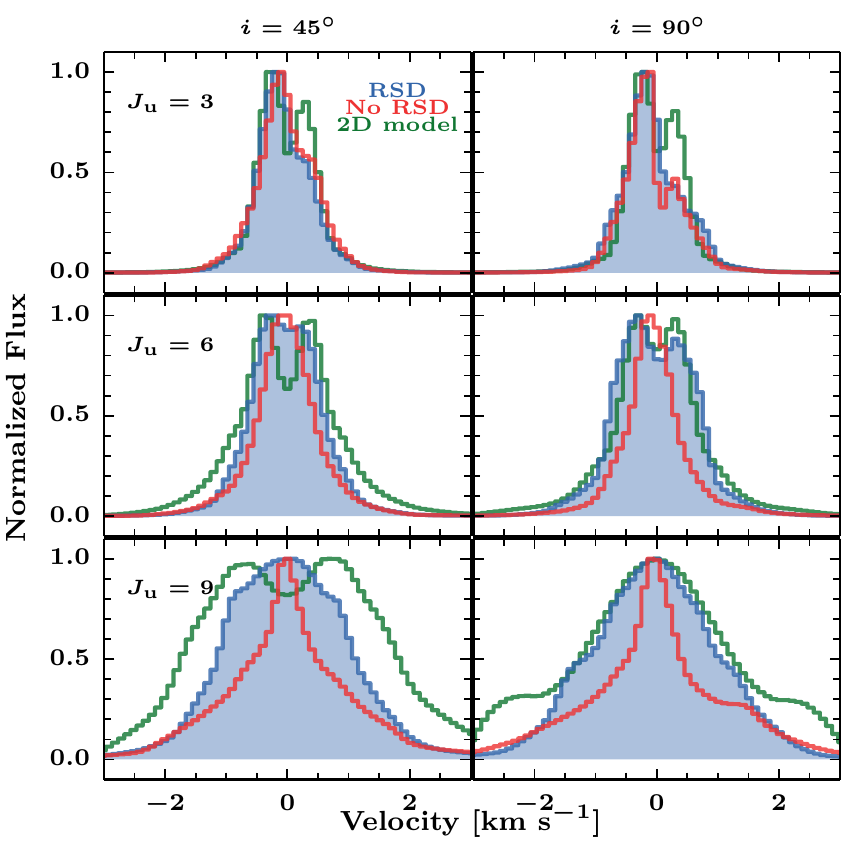}
\caption{\ceo\ line profiles for the different simulations viewed at 
$i = 45^{\circ}$ and $i=90^{\circ}$ in $15\arcsec$ beam. }
\label{fig:inclconv}
\end{figure}

The simulated lines viewed face-on may not pick up all of the 
dynamics of the system.  Figure~\ref{fig:inclconv} shows how the line 
profiles change with inclination for the three different simulations 
within a $15\arcsec$ beam.  The lines become broader with increasing 
inclination as they readily pick up the different velocity 
components.  The $J_{\rm u} = 3$ lines exhibit inverse P-Cygni line 
profiles that are associated with infalling gas.  Meanwhile, the 
higher $J$ transitions show more structured line profiles compared 
with the systems viewed face-on.  The \ceo\ 6--5 lines are 
double-peaked in both cases of RSD formation while it is 
single-peaked for the pseudo-disk model.  On the other hand, the 9--8 
line is significantly broader than the low-$J$ lines reflecting the 
complexity of the dynamics of the warm dense gas.


\subsubsection{Origin of the line broadening}\label{sect:oribroad}

To investigate the source of the line broadening, a set of molecular 
lines are simulated with zero radial velocity ($\upsilon_{\rm r} = 0$ 
\kms) and another with zero azimuthal velocity ($\upsilon_{\phi} = 0$ 
\kms).  For the MHD simulation with RSD, the $FWHM$ value of the 
\ceo\ 9--8 line decreases to $< 0.5$ \kms\ at all inclinations 
without any azimuthal velocity component.  Such a decrease is not 
dramatic for face-on orientation, however it is more than a factor of 
3 for intermediate ($i \sim 45^{\circ}$) and high inclination ($i > 
75^{\circ}$) cases.  On the other hand, in the case of pseudo-disk 
formation, both radial and azimuthal velocities are of equal 
importance.  The origin of the line broadening therefore depends on 
whether or not an RSD is forming.  If an RSD is indeed forming, the 
\ceo\ 9--8 is broadened by rotational motions at moderate and high 
inclinations; at low inclinations, infall dominates the broadening.


\subsubsection{Line widths and comparison with observation}

\begin{table}
 \centering
 \caption{$FWHM$ ($FWZI$ as defined at 10\% of the peak emission) in 
\kms\ of the \tco\ and \ceo\ lines within a 20$\arcsec$ beam for the 
MHD simulations viewed at $i = 45^{\circ}$. }
 \label{tbl:fwhms0}
 \begin{tabular}{l c c | c c}\toprule\hline
  & \multicolumn{2}{c}{\tco} & \multicolumn{2}{c}{\ceo} \\ 
$J$ & RSD & No RSD & RSD & No RSD \\ \hline 
3--2 & 0.8 (1.8) & 1.1 (2.2) & 0.9 (1.8) & 0.9 (1.9) \\
6--5 & 1.1 (2.3) & 1.0 (2.2) & 1.3 (2.3) & 0.9 (2.2) \\
9--8 & 2.1 (3.7) & 1.2 (3.9) & 2.3 (4.0) & 1.0 (3.9) \\ \hline
 \end{tabular}
\end{table}

\begin{figure}[htbp]
\centering
\includegraphics{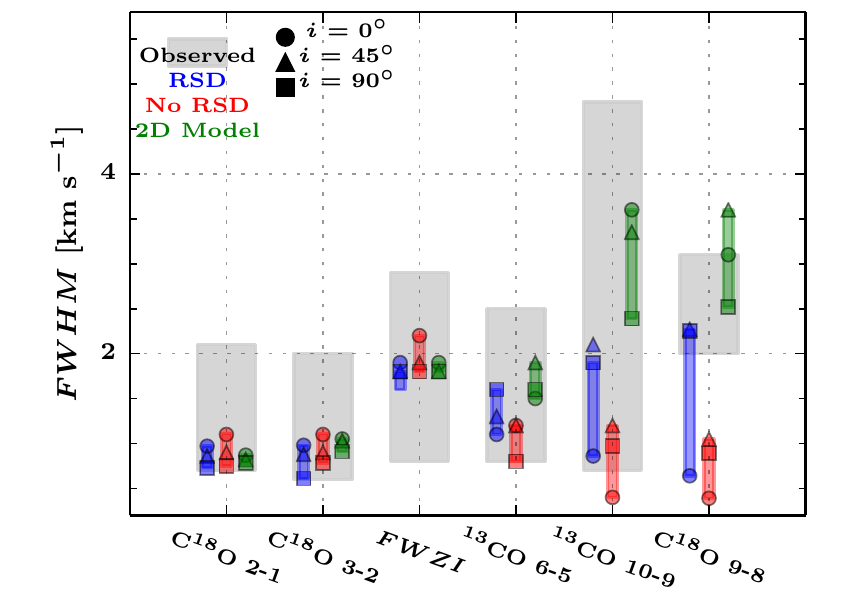}
\caption{$FWHM$ values of the observed (gray shaded region) and 
simulated (symbols) rotational lines of \tco\ and \ceo\ as indicated 
in a 20$\arcsec$ beam.  The observed values are taken from 
\citet{jorgensen02}, \citet{vankempen09c}, and 
\citet{sanjose-garcia13}.  The $FWHM$ of synthetic lines are 
indicated at specific inclinations as indicated by the symbols: 
circles for $0^{\circ}$, triangles for $45^{\circ}$, and squares for 
$90^{\circ}$.  The comparison of the $FWZI$ values are specifically 
for the \ceo\ 3--2 line.  The different colors represent the 
different type of simulations: 3D RSD (RSD), 3D pseudo-disk (No RSD), 
and 2D semi-analytical model (2D Model). }
\label{fig:linewidthspread}
\end{figure}


Molecular line observations are typically characterized by their peak 
flux densities (or intensities), $FWHM$, and integrated line flux 
densities.  While the peak flux densities and integrated line fluxes 
depend on the adopted physical and chemical structure, their $FWHM$ 
should reflect the general kinematics that are present in the system.  
In this paper, we focus on the comparison of $FWHM$ and full-width at 
zero intensity ($FWZI$) determined at 10\% of the peak with 
observations.  A 10\% cut-off is chosen since most single-dish 
observations do not reach higher signal-to-noise, especially for the
higher-$J$ lines.

These values are calculated for \tco\ and \ceo\ lines of the 
different models from the convolved image cubes.  The $FWHM$ and 
$FWZI$ within a 20$\arcsec$ beam are listed in Table~\ref{tbl:fwhms0} 
at moderate inclination ($i = 45^{\circ}$) comparing the two 3D MHD 
simulations.  Within such a large beam, their values for \tco\ and 
\ceo\ are similar.  The $FWHM$ values do not necessarily increase 
between $J_{\rm u}=$3 and 6, in contrast with the $FWZI$ values (see 
Fig.~\ref{fig:linewidthspread}).  This is expected since the wing 
emissions are much lower than the peak because the emitting region is 
much smaller than the beam (beam dilution).

Considering all inclinations ($0^{\circ}$, $45^{\circ}$, and 
$90^{\circ}$), the $FWHM$ values of the low-$J$ CO lines are similar 
between the 3D simulations and 2D semi-analytical models.  On the 
other hand, the line widths of the high-$J$ lines differ 
significantly.  This is expected since most of the 2--1 and 3--2 
emission originates from the large-scale envelope, which is similar 
in terms of kinematics in the three simulations.  However, the 
high-$J$ lines originate from the warm inner regions in which the 
three simulations show different velocity structures.

Figure~\ref{fig:linewidthspread} presents the comparison between the 
simulated and observed lines.  The observed line widths toward a 
sample of Class 0 low-mass YSOs are taken from \citet{jorgensen02}, 
\citet{vankempen09c}, and \citet[][$FWHM_{\rm N}$ for the narrow 
component]{sanjose-garcia13}.  Sources with known confusion in their 
line profiles from other nearby sources have been excluded.  It is 
clear that the observed line widths of the low-$J$ lines ($J_{\rm 
u}=$ 2 and 3) are significantly greater than the model simulations, 
by a factor of 2.  Since these lines probe the large-scale quiescent 
envelope, the discrepancy between the predicted and observed line 
widths suggests that the large-scale envelope is turbulent with $FWHM 
\approx$ 1 \kms\ or a Doppler $b$ of $\sim 0.4$ \kms, i.e., more than 
what is included in the current simulations that could be due to an 
interaction with fast outflow, at least in part. The comparison of 
the $FWZI$ values also supports this conclusion.

For the dense and warm gas probed by the higher $J_{\rm u} \ge 6$ 
lines, the predicted line widths are within the observed range 
for the models that do form an RSD.  Moreover, the pseudo-disk (No 
RSD) simulations predict much smaller line widths than those 
observed.  This line originates from the inner warm parts that are 
rotationally supported.  Our treatment of the inner radius does not 
affect this since an evacuated cavity is absent in the case of RSD 
formation.  In fact, if an outflow cavity is allowed to form, the 3D 
RSD predicted line widths would shift upward.  Thus, the line width of 
this particular line reflects the true kinematics in the inner 
regions.  This comparison may suggest that the large observed line 
widths of \ceo\ 9--8 line indicate that the kinematics of the warm 
inner envelope are similar to that of a model that forms an RSD.  
Alternatively, a turbulence with $FWHM=$ 2 \kms\ would be required in 
the inner parts if RSDs are absent.


\section{Discussion}\label{sec:dis}


\subsection{Variations with viewing angles}

\begin{figure*}[htbp]
\centering
\includegraphics{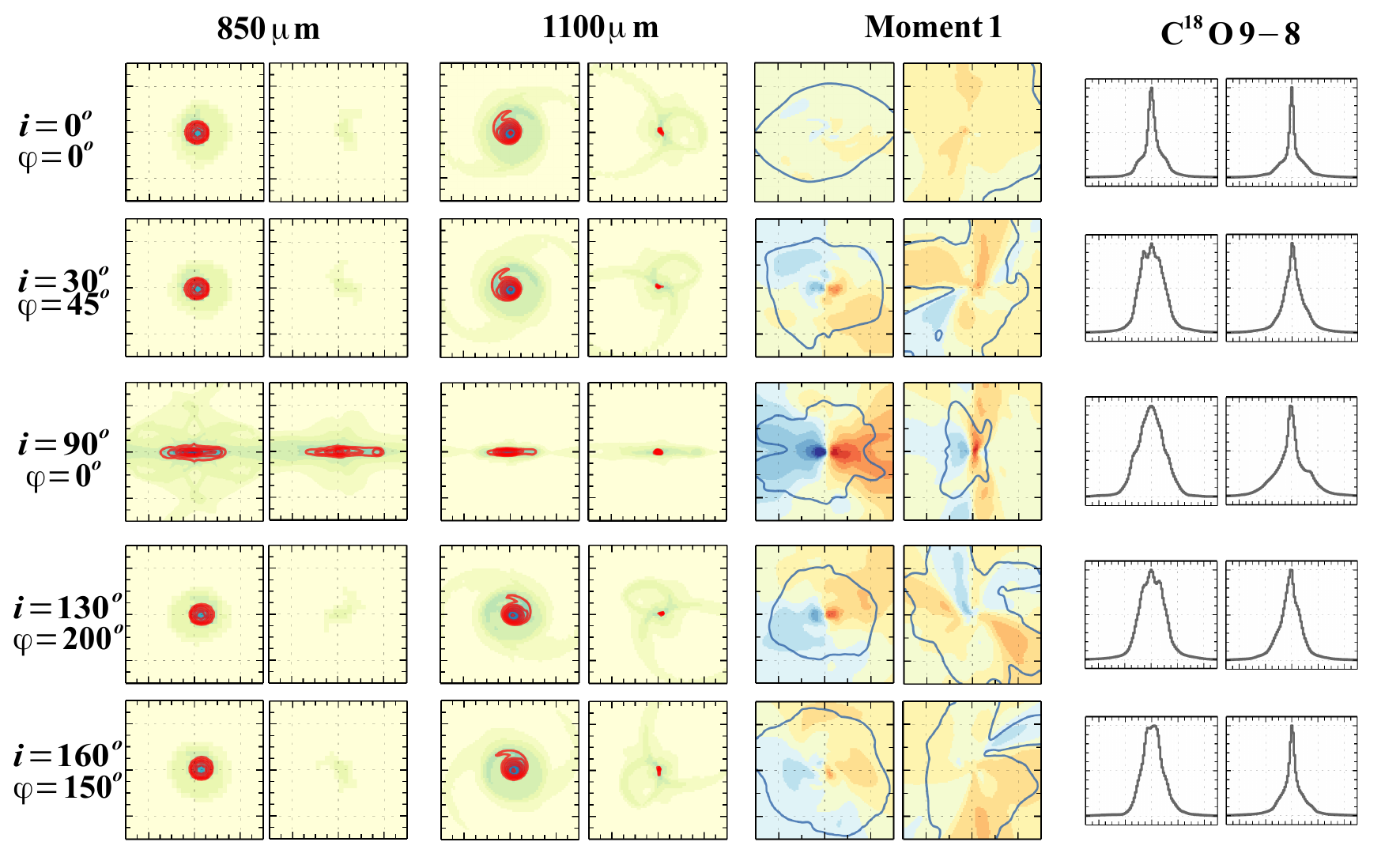}
\caption{Comparison of synthetic 850 $\mu$m, 1100 $\mu$m, \ceo\ 2--1 
moment one maps and normalized \ceo\ 9--8 spectra for the 3D RSD 
({\it left}) and pseudo-disk ({\it right}) MHD simulations viewed at 
a few orientations.  The images and image cubes are convolved with a 
$0.1\arcsec$ beam.  The maps sizes are 5$\arcsec$ across similar to 
Fig.~\ref{fig:convimage}.  The color scales and line contours are 
defined in Figs.~\ref{fig:convimage} and \ref{fig:moment13265}.  
The normalized \ceo\ 9--8 spectra refer to a 15$\arcsec$ beam on 
the same scale as Fig.~\ref{fig:inclconv}.  The inclination 
($i$) and azimuthal angle ($\phi$, rotation angle of the 
observer around the $z$ axis) are representative only and not exact.}
\label{fig:viewangle}
\end{figure*}

This paper presents synthetic continuum maps and CO isotopolog lines 
from 3D MHD simulations and 2D semi-analytical disk formation models 
out of collapsing rotating envelopes.  The aim is to present 
signatures that can differentiate between embedded rotationally 
supported disks (RSDs) and a pseudo-disk.  Thus far, we have analyzed 
the continuum and synthetic molecular lines with respect to one 
viewing angle in the azimuthal direction with $\phi = 0^{\circ}$.  As 
shown previously by \citet{rowansmith12}, the line profiles may 
change with different viewing angles since the collapse process is 
not spherically symmetric.

In order to look at the general trend with viewing angles, synthetic 
images at 4 different inclinations ($i$) from 15$^{\circ}$ to 
150$^{\circ}$ and 8 azimuthal angles from 8$^{\circ}$ to 
315$^{\circ}$ are generated.  We concentrate the analysis on 850 
$\mu$m, 1100 $\mu$m, \ceo, and \cso\ images.  
Figure~\ref{fig:viewangle} presents the continuum and molecular lines 
at a few viewing angles for the two 3D MHD simulations.  The 3D RSD 
formation predicts an observable spiral feature at near face-on 
($i\approx 0^\circ$) with high dynamic range (1000).  As mentioned 
earlier, this is due to the infalling material on its way to the disk.

In terms of kinematical signatures, the moment one maps of the \ceo\ 
2--1 line are compared for the two cases.  Both the RSD and 
pseudo-disk simulations predict a coherent velocity gradient in the 
inner 300 AU.  However, the velocity gradient is more robust if an 
RSD is forming.  The pseudo-disk simulation shows a velocity gradient 
in the moment one map that is not necessarily corresponding to 
rotation (see Fig.~\ref{fig:phys2d} and Section~\ref{sec:mommap}).  
Such a direction corresponds to the streamlines of the infalling 
material from the large-scale envelope to the central star.  At high 
inclinations ($i > 75^{\circ}$), this direction of the velocity 
gradient shifts to east-west direction similar to the RSD simulation, 
because the moment one map is dominated by the rotational motions 
which are in the east-west direction.

To assess the general predictions for the large-scale envelope, \ceo\ 
9--8 spectra within a 15$\arcsec$ beam are compared.  The low-$J$ 
lines ($J_{\rm u}=$ 2 and 3) exhibit inverse P-Cygni profiles 
indicating infalling material in both RSD and pseudo-disk 
simulations. For the high-$J$ lines, the 3D RSD simulation predicts a 
broader line than the pseudo-disk simulation in most orientations 
consistent with Fig.~\ref{fig:linewidthspread}.

In summary, the results that are presented in previous sections are 
robust and do not depend on the viewing angles.  A pseudo-disk shows 
more distinct features in the kinematics in the moment one maps 
that are different from an RSD.  A coherent blue- to red-shifted 
velocity gradient in the inner 1 arcseconds aligned with the major 
axis is most likely a signature of an RSD or on-going RSD formation.  
Although these lines are simulated assuming LTE conditions, non-LTE 
effects generally decrease the strength of the emission but do not 
alter the results of the kinematics derived from the molecular limes.


\subsection{Masses and sizes}

\begin{figure*}[htbp]
\centering
\begin{tabular}{cc}
\includegraphics{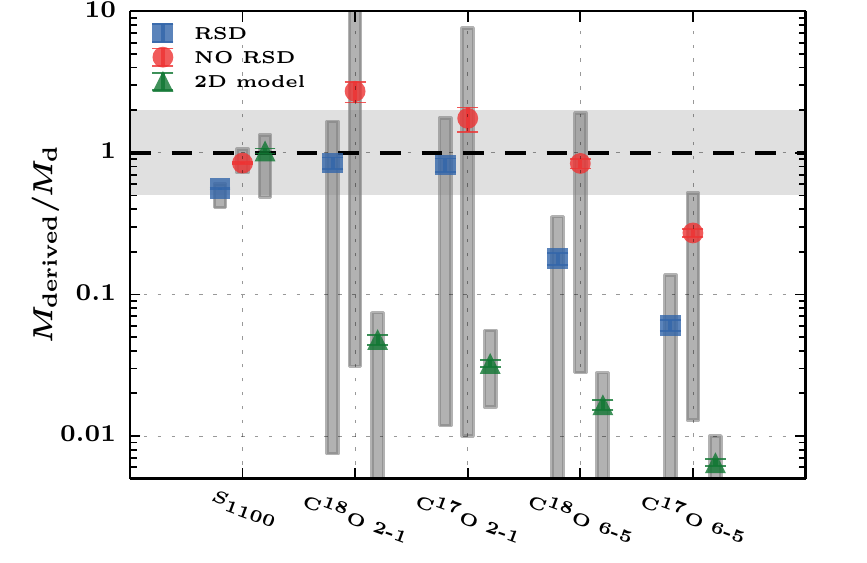}
 &
\includegraphics{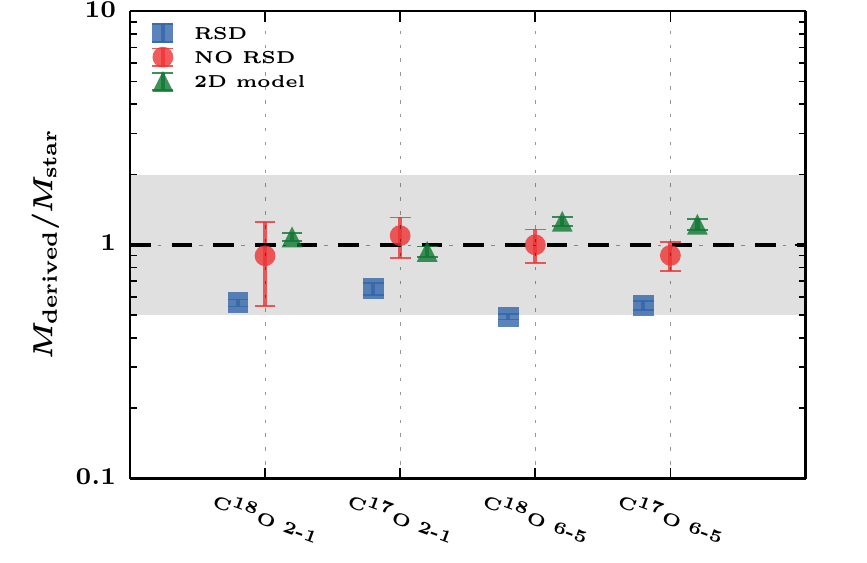} 
\end{tabular}
\caption{{\it Left: } Mean deviation of disk masses (gas + dust) 
derived from 1100 $\mu$m continuum and molecular lines fluxes from 
the true disk mass extracted from the simulations.  The gray 
bars indicate the full range of values for the different 
orientations.  {\it Right: } Mean deviation of stellar masses measured 
from molecular lines from the true stellar mass.  The colors indicate 
the different simulations: 3D MHD simulations in blue, 3D MHD 
pseudo-disk simulation in red, and 2D semi-analytical model in green.  
The error bars indicate the standard error of the mean.  The shaded 
region shows the area within a factor of 2. }
\label{fig:massderiv}
\end{figure*}

\subsubsection{Analysis with continuum}

To test models of star and disk formation \citep[e.g.,][]{hueso05}, 
the flow of mass from envelope to disk to star with time needs to be 
known. In practice, this means that the masses of the central 
(proto-)star and of its envelope-disk system need to be inferred from 
observations and models at different evolutionary stages.  A 
pragmatic way to `extract' the envelope and disk masses is to compare 
the continuum fluxes within a large beam with those found at `small' 
scales.  Here, 5$\arcsec$ area is taken as the `small' scale  
corresponding to 50k$\lambda$ at a distance of 140 pc as probed 
previously with the SMA.  Specifically, \citet{prosac09}\footnote{ 
Note that the factor $\left ( \frac{1.1}{0.85} \right )^{\alpha}$ is 
omitted since the analysis is performed on the same wavelength.}   
gives the following formulae: 
\begin{eqnarray}
 S_{\rm 5 \arcsec} & = & S_{\rm disk} + f_{\rm env} \times S_{\rm env} 
\\
 S_{\rm 15 \arcsec} & = & S_{\rm disk} + S_{\rm env},
 \end{eqnarray}
where $f_{\rm env}$ is the envelope contribution at `small' scales, 
$S_{\rm 5\arcsec}$ is the flux within a 5$\arcsec$ area, $S_{\rm 15 
\arcsec}$ is the flux within a $15\arcsec$ beam, $S_{\rm env}$ is the 
total envelope flux, and $S_{\rm disk}$ is the total disk flux.  The 
set of equations above roughly separates the envelope from the 
disk in terms of continuum flux contribution.   There are 
two issues that need to be addressed in utilizing the equations 
above: the envelope fraction ($f_{\rm env}$) and the `small' scale
flux.

The first issue is presented in \S 4.1 of \citet{prosac09}.  They 
calculated the envelope contribution as determined from $\rho_{\rm 
dust} \propto r^{-1.5}$ spherical envelopes.  This yields a maximum 
contribution of 8\% at 850 $\mu$m within a scale of 5$\arcsec$ 
compared with that at $15\arcsec$.  The contribution increases with 
increasing power-law exponent (16\% for $\rho \propto r^{-1.8}$).  
Note that \citet{prosac09} computed the relative contribution between 
$S_{\rm 50 \ k\lambda}$ at 1.1 mm and $15\arcsec$ in 850 $\mu$m.  In 
order to compute the contributions within the same wavelengths, these 
values are scaled assuming $F_{\nu} \propto \nu^{2.5}$.  In addition, 
for a given 2D embedded disk model, the projected spherical envelope 
model is one with an increasing exponent with inclination (i.e., 
$\rho \propto r^{-1.8}$ as $i \rightarrow 90^{\circ}$).  In other 
words, for an embedded YSO viewed at face-on, the best representative 
spherical model is one with a rather flat density profile.  As a 
result, the envelope contribution from  an embedded YSO on small 
scales changes with inclination.

Another issue is the `small' scale flux at which to determine the 
envelope contribution.  If the size is too small (i.e., 1$\arcsec$), 
the envelope contribution is naturally negligible and it is easy to 
extract the disk flux.  However, the disk flux can be underestimated 
within such a small scale.  Furthermore, the flux that is measured at 
each pixel is {\it always} a combination of the envelope and the 
disk.  Thus, it is more intuitive to calculate the envelope's 
contribution at scales larger than the disk ($\sim 5\arcsec$, 700 AU 
at 140 pc) in order to obtain all of the disk's flux.

For this purpose, following the method in the literature, the 
envelope and disk masses are determined from comparing the fluxes 
within a $15\arcsec$ beam and a $5\arcsec$ area.  This procedure is 
performed on the images rendered at the 32 different viewing angles 
as described in the previous paragraphs.  A typical constant envelope 
fraction of 8\% at 850 $\mu$m (2\% at 1100 $\mu$m) is adopted to 
see how well such a simple estimation based on spherically symmetric 
envelope model can extract the disk properties of the 2D and 3D 
simulations.

Once the disk and envelope fluxes at 850 $\mu$m have been obtained, 
the conversion to dust mass follows \citet{prosac09}, Eq.~(1), which 
takes into account the fact that there is a distribution of 
temperatures in the envelope set by the luminosity of the protostar.  
The envelope mass therefore scales both with distance and bolometric 
luminosity ($L_{\rm bol}$).  The bolometric luminosities are 
calculated by constructing the spectral energy distribution at the 
same viewing angles as the synthetic images.  For the disk, a dust 
temperature of 30 K is adopted to calculate its mass as it is done in 
observations.  These values are compared to the masses of different 
components in the simulations (Table~\ref{tbl:params}).  The disk mass 
in the simulation is defined by the velocity structure as described 
in Section~\ref{sec:rsd}.  The left-over material within the 
computational box is the envelope mass.

Figure~\ref{fig:massderiv} presents the ratio of disk masses derived 
from synthetic continuum observations to the true masses tabulated in 
Table~\ref{tbl:params}.  In the case of the pseudo-disk,  the disk 
mass is taken to be the mass of the region with number densities 
$n_{\rm H_2} > 10^{7.5}$ cm$^{-3}$.  The masses inferred from the 
continuum fluxes are within a factor of 2 of the true disk mass.  
There is a little difference in the disk mass between 1100 $\mu$m and 
850 $\mu$m for both 3D RSD and 2D semi-analytical model.  Envelope 
masses agree if scaled to the same $15\arcsec$ beam.

The spread in the inferred disk masses is greater in the 2D model 
than in the 3D simulation.  The lower end of the spread is occupied 
by simulations at high inclinations ($i > 75^{\circ}$).  In such an 
orientation, the continuum optical depth even at 1100 $\mu$m is high 
since the disk is viewed edge-on and, consequently, a large part of 
the disk does not contribute to the observable emission.   The high 
end of the spread is when the system is viewed almost face-on where 
the inner flattened inner envelope deviates from the spherically 
symmetric assumption.  Finally, the equations above underestimate the 
disk mass from the 3D simulations because a $T_{\rm dust} = 30$ K was 
assumed to obtain the disk mass whereas the temperature within the 
disk is $< 30$ K at large radii.  A lower temperature $\sim 10$ 
K would be best for the 3D RSD simulation while it is inclination 
dependent for the 2D model case.

The results suggest that disk masses can be well estimated from 
the continuum flux even for the pseudodisk.  However, there is a 
clear difference in how the disk is defined in all three different 
cases.  In the 3D and 2D RSD, the disk mass is defined by its 
velocity structure such that it is indeed rotationally supported 
while the pseudodisk is defined by density.  Hence, in the latter 
case, the difference between the disk and the infalling envelope 
is not well defined.

In summary, we find that the disk masses as inferred from continuum 
emission in the 3D simulations and 2D semi-analytical model are 
within a factor of 2 of the true value.  This factor-of-2 is due to 
the viewing angle and the dust temperature within the disk is lower 
than the assumed 30 K.  However, we find that the pseudo-disk also 
indicate similar emission at such scales \citep[see 
also,][]{chiang08}. In this case, the disk does not corresponds to a 
Keplerian disk.


\subsubsection{Disk radii and stellar masses}

Analysis of the continuum toward a large sample of sources yields 
disk and envelope masses.  In order to fully test the evolutionary 
models, the stellar masses need to be derived from the kinematics of 
molecular lines.  In addition, the disk radii can be determined from 
the velocity profiles (see Section~\ref{sec:velprof}) and can also 
serve as tests for the evolutionary models since disk radius is 
expected to increase with evolutionary state but to also depend on 
the initial angular momentum of the core (see Fig.~13 in 
\citealt{harsono14}).  The extent of the RSD is determined from the 
break in the peak-position velocity diagrams (see 
Fig.~\ref{fig:velprof1}) for the simulated \ceo\ 2--1, \ceo\ 6--5, 
\cso\ 2--1 and \cso\ 6--5 lines.  For the 2D semi-analytical models, 
the average radius from the 4 lines is $64 \pm 8$ AU as derived from 
the break.  For the 3D MHD RSD simulation, the average radius is $230 
\pm 100$ AU.  Thus, the inferred drop-off gives a good estimate of 
the extent of small disks, while a larger disk has a larger error bar 
associated to it.

Disk masses can also be calculated from the integrated flux densities 
($\int S_{\nu} d\upsilon$ in Jy \kms) of molecular lines.  In this 
method, the integrated flux densities are extracted within the region 
defined by the radius in the previous paragraph.  The mass calculation 
is given by \citep{scoville86, momose98, hogerheijde98}:
\begin{eqnarray}
 M_{\rm gas} & = & 5.9 \times 10^{6} \frac{Q(T_{\rm ex})}{ g_{\rm u} 
A_{\rm ul} \exp{\left (-E_{\rm u} / k_{\rm B} T_{\rm ex} \right ) }  
} \frac{\mu_{\rm H_2} m_{\rm H_2} }{[X/{\rm H_2}]} \\ \nonumber & & 
\frac{\tau_{\rm L}}{1 - \exp^{-\tau_{\rm L}}} \left ( \frac{d}{\rm 
140 pc} \right )^2 \int S_{\upsilon} d\upsilon  \  M_{\odot},
\end{eqnarray}
where $A_{\rm ul}$ is the Einstein A coefficient of the transition, 
$g_{\rm u}$ is the degeneracy of the upper level, $E_{\rm u}$ is the 
upper level energy, $Q(T_{\rm ex})$ is the partition function at an 
excitation temperature $T_{\rm ex}=40$ K, $\mu_{\rm H_2}$ is the mean 
weight of the gas, $[X/ \rm H_2 ]$ is the abundance of molecule $X$ 
with respect to $\rm H_2$, $\tau_{\rm L}$ is the line optical depth, 
and $k_{\rm B}$ is the Boltzmann constant. A higher excitation 
temperature for the gas is adopted with respect to the dust 
temperature following the rotational temperature derived from the 
observed C$^{18}$O lines \citep{yildiz13}.  The derived masses do not 
strongly dependent on the adopted excitation temperature.  A 
constant line optical depth of 0.5 for the \ceo\ and 0.3 for the 
\cso\ lines is used, which characterizes the average optical depth 
over all velocities.  The crucial assumption here is the CO 
abundance, $[X/ \rm H_2 ]$, which is taken to be $10^{-4}$ similar to 
that in Section~\ref{sec:radtrans}. Since the CO abundance will be 
affected by freeze-out in the cooler parts of the disk, this 
assumption provides a lower limit to the disk mass.

Figure~\ref{fig:massderiv} presents the masses obtained from the 
integrated line flux densities.  They show a much larger scatter than 
the masses obtained from the dust continuum flux.  The symbols 
indicate the values appropriate for masses derived at moderate 
inclinations.   For the case of RSD formation, the low-end of the 
spread is due to near face-on orientation ($30^{\circ} > i > 
150^{\circ}$) in which the obtained disk sizes are half of the true 
RSD and, consequently, the disk mass is lower than average.  In the 
2D semi-analytical model case, the physical structure of the inner 
envelope plays a role.  The conversion between integrated flux to 
mass also assumes that the emission at all velocities is dominated by 
the disk, which is only true for emission at the wings (high 
velocities $\gtrsim 1$ km s$^{-1}$).  However, in the 2D 
semi-analytical model, the inner envelope is highly flattened 
compared to that of the 3D MHD simulation such that high density gas 
($n_{\rm H_2} > 10^6$ cm$^{-3}$) surrounds the RSD (see 
Fig.~\ref{fig:phys2d}).  So, the disk does not dominate the 
integrated flux density at this particular time \citep[see Fig.~8 
of][]{harsono13} and, consequently, the integrated flux density is 
not correlated with its mass.  Therefore, a good estimate of the 
disk's molecular mass can only be obtained if the system is oriented 
at moderate inclinations and the RSD is assumed to dominate the 
molecular emission at all velocities.

The disk masses derived from the 2--1 line are generally higher than 
that from the 6--5 line.  Since the RSD size obtained from the 6--5 
emission is smaller than that from the 2--1 line (e.g., 200 vs 100 
AU), the integrated line flux density is extracted from a smaller 
region, which results in a much lower mass.  The size of the RSD is 
smaller in the 6--5 emission because the emission arises from the 
dense warm regions in the vicinity of the protostar and, thus, its 
mass.  The masses obtained from the integrated 2--1 line are 
better estimates for the true disk mass.

The 3D MHD and 2D semi-analytical RSD formation predict the same 
behaviour in terms of disk masses in both \cso\ and \ceo\ lines.  The 
difference of the masses obtained in the 2--1 and the 6--5 lines are  
also similar for the two isotopologs.  The 2D semi-analytical model 
predicts a smaller spread than the 3D RSD model in the disk masses 
obtained from the \cso\ 2--1 line since the high density region is 
more compact in the 2D case and, therefore, the \cso\ 2--1 integrated 
flux density obtained in the 2D model does not vary as much as in the 
3D simulation.  The comparison between the 2D semi-analytical model 
and 3D MHD shows that the reliability of the observables to trace the 
true masses depends on the physical structure of the inner envelope.

The stellar masses obtained from the PPV method are shown in 
Fig.~\ref{fig:massderiv}.  They are typically within a factor of 2 of 
the true stellar masses.  The best-fit values for synthetic molecular 
lines viewed at $i < 15^{\circ}$ tend to be more than twice the true 
stellar mass.  This is due to the difficulties in obtaining the peak 
positions at high velocities since they are most likely to be below 
the noise level.  The 2D semi-analytical model indicates better 
agreement with true stellar masses than those of the 3D model because 
their velocity structures are different.  The kinematics of the 
envelope is the same at all viewing angles in 2D.  However, there are 
lines of sights that pick up significant sub-Keplerian gas infalling 
onto the disk in 3D as shown in Fig.~\ref{fig:simvelprofs}.  
Therefore, stellar masses as obtained from spatially and spectrally 
resolved molecular observations are good estimates of the true 
stellar mass within a factor of 2, which is smaller than the 
uncertainty in the inclination.

From the analysis of the synthetic observations, spatially 
($\lesssim 0.1\arcsec$ at 140 pc) and spectrally ($\leq 0.1 \ {\rm km 
\ s^{-1}}$) resolved optically thin molecular lines at two energy 
levels (e.g., 2--1 and 6--5) are required to understand the physical 
structure of the inner envelope.  The disk masses obtained from the 
two lines are similar in the pseudodisk case while they can differ by 
an order of magnitude in RSD case.  The simple PPV analysis directly 
from the data can already differentiate between the RSD and a 
pseudodisk.  However, sophisticated modelling tools are needed to 
infer the physical structure of the disk.


\section{Summary and conclusions}\label{sec:sum}

We have presented the observables in continuum and molecular lines 
for two 3D MHD simulations of \citet{zyli13} and 2D semi-analytical 
models of collapse and disk formation.  Snapshots of two different 
MHD simulations of a relatively weakly magnetized core ($B_0 = 11$ 
$\mu$G) at the same time after the onset of collapse are used.  One 
simulation has an initial magnetic field axis aligned with the 
rotation axis in which a rotational supported disk (RSD) does not 
form, however a pseudo-disk and outflowing gas are present.  The other 
MHD simulation starts with the magnetic field axis oriented 
perpendicular to the rotation axis, which results in the formation of 
an RSD (see Figs.~\ref{fig:phys2d} and~\ref{fig:simvelprofs}).  These 
simulations explore the two extremes of the magnetic field 
orientation. The synthetic observables are then compared to a 2D 
semi-analytical model without magnetic field \citep{visser09} with 
similar initial conditions (1 $M_{\odot}$ and $\Omega_0 = 10^{-13}$ 
Hz).  Accurate dust temperatures are calculated using the 3D continuum 
radiative transfer tool RADMC3D with the same dust opacities and 
central temperature for all three models.  Continuum images and 
thermalized CO molecular lines are produced using the same radiative 
transfer code and method.  Freeze-out of CO onto dust grains is 
included.  This paper focuses on presenting similarities and 
differences in the predicted observables.  The main results and 
conclusions are as follows.

\begin{itemize}

\item Synthetic continuum images of the two MHD simulations and 2D 
    semi-analytical model indicate that a spatial resolution of 14 AU 
    and high dynamic range (1000) are required to differentiate 
    between disk formation scenarios.  Furthermore, the features that 
    are present during the collapse are more easily observed in the 
    450 $\mu$m continuum images than at longer wavelengths.  It is 
    difficult to test disk formation models toward highly inclined 
    systems using continuum data since both RSD and pseudo-disk 
    formation show similar elongated features.

\item The kinematical structures as revealed by the moment one 
    maps of the synthetic molecular lines show a coherent blue- to     
    red-shifted velocity gradient for both RSD models and for the 
    pseudo-disk in the inner $\sim 300$ AU.  However, the pseudo-disk 
    shows a velocity gradient in north-south direction while the RSD 
    shows an east-west gradient similar to the orientation of the 
    flattened structure in the model.  Moreover, the RSD formation in 
    both 3D and 2D exhibits skewness in their moment one maps caused 
    by the infalling rotating envelope component on larger scales 
    which is absent in the case of pseudo-disk.  The velocity 
    gradient in the case of the pseudo-disk is a nearly straight 
    line from the star to the large-scale structure since it is 
    tracing the streams of material directly from the envelope onto 
    the star.  Thus, one can readily mistake a pseudo-disk with an 
    RSD unless one performs additional analysis such as the peak 
    position diagrams.

\item Position-velocity (PV) diagrams constructed from  \ceo\ and 
    \cso\ image cubes predict rotational signatures in both
    pseudo-disk and RSD formation, seen most prominently in \cso\ 
    data.  This is due to the strength of rotation in the inner 
    regions for both simulations.  A combination of \ceo\ and \cso\ 
    lines is required to disentangle the RSD from a pseudo-disk.  The 
    signatures of infalling material in the pseudo-disk simulation 
    are stronger in the \ceo\ lines.  The velocity structure 
    constructed from the peak positions ({\it peak-PV diagrams}) are 
    used to differentiate between the pseudo-disk and the RSD.  
    Velocity structures described by $\upsilon \propto r^{-0.5}$ 
    and $\upsilon \propto r^{-1}$ are present in both cases of RSD 
    formation whereas a flatter velocity profile is seen in the 
    pseudo-disk case.  We find that this conclusion is robust for 
    different inclinations and rotations.

\item The image cubes are convolved with large beams ($\ge 9\arcsec$) 
    to simulate single-dish observations probing the large-scale 
    envelope.  The \ceo\ 2--1 and 3--2 line widths are similar 
    between the three simulations with $FWHM \sim 1$ \kms or Doppler 
    $b$ of $0.4$ \kms.  This is due to the fact that the emitting 
    regions in the large-scale envelopes are similar.  The observed 
    $FWHM$ values are larger than in those predicted from the 
    simulations by a factor of 2.  This suggests that the large-scale 
    envelopes of low-mass embedded YSOs are significantly more 
    turbulent than these models, which may be due to an interaction 
    with a fast outflow \citep{sanjose-garcia13}.

\item The comparison of the high-$J$ lines (\tco\ 6--5, \tco\ 10--9,  
    and \ceo\ 9--8) indicates that the current simulations with RSD  
    formation can reproduce the observed line widths solely due to 
    the rotation + infall motions.  On the other hand, the predicted 
    line widths from the pseudo-disk are significantly smaller than 
    the observed values.  Thus, the mechanism(s) that are responsible 
    for broadening the 6--5 and 9--8 lines depend on the whether or 
    not an RSD is present.  If no RSD is  present, the observations 
    would imply an increasing level of turbulence with decreasing 
    radii; if an RSD is present, turbulence would not be needed,
    although it can still be present in the disk at some level.

\item Masses derived from continuum and molecular lines from 
    simulations and analyzed in the same way as observations depend 
    on the physical structure at small-scales.  The disk masses 
    obtained from the continuum flux in small beams (a few $\arcsec$) 
    are generally in agreement with the true disk mass.  Disk masses 
    obtained from the integrated molecular line flux depend strongly 
    on the physical structure.  If the disk is small and cold with 
    respect to the flattened inner envelope, inferred masses can be 
    smaller by an order of magnitude from the true disk mass since 
    the disk does not contribute to the integrated line flux density. 
    However, if the disk is large and warm enough to prevent 
    significant freeze-out, the mass obtained from the low-$J$ lines 
    can give a good estimate of the true RSD mass provided that the 
    system is inclined toward us.  Both the stellar masses and 
    disk-to-envelope mass ratios are within a factor of 2 of the true 
    masses.  The presence of the RSD cannot be determined solely from  
    continuum data alone, however the continuum flux provides a good  
    estimate on the mass at small scales.  Multiple spectrally and 
    spatially resolved molecular line observations are needed to 
    confirm the presence of RSD.
    
\end{itemize}


\section*{Acknowledgments}

We thank Atilla Juh{\'a}sz for providing scripts for generating and 
analyzing RADMC3D input and output files.  We also thank Kees 
Dullemond for providing RADMC3D.  We are grateful to Michiel 
Hogerheijde, Floris van der Tak, and Lee Hartmann for commenting on 
the manuscript.  We thank the anonymous referee for the useful 
comments that have improved this manuscript.  This work is supported 
by the Netherlands Research School for Astronomy (NOVA).  
Astrochemistry in Leiden is supported by the Netherlands Research 
School for Astronomy (NOVA), by a Royal Netherlands Academy of Arts 
and Sciences (KNAW) professor prize, and by the European Union A-ERC 
grant 291141 CHEMPLAN.  ZYL is supported in part by NASA 10AH30G, 
NNX14B38G, and NSF AST1313083.


\bibliographystyle{aa}
\bibliography{../../biblio.bib}


\end{document}